\documentclass[aps,prc,10pt,showpacs,showkeys,twocolumn,superscriptaddress,groupedaddress]{revtex4-1}
\usepackage[latin1]{inputenc}
\usepackage[english]{babel}
\usepackage{amsmath}
\usepackage{amssymb}
\usepackage{amscd}
\usepackage{eucal}
\usepackage{color}
\usepackage{bm}
\usepackage{braket}
\usepackage{dsfont}
\usepackage{float}
\usepackage{graphicx}
\usepackage[makeroom]{cancel}
\usepackage{bbm}
\usepackage[capitalize]{cleveref}
\bibstyle{asprev}

\newcommand{\nmax}{N_{\rm max}}
\newcommand{\hb}{\hbar\Omega}
\def\id{\mathds{1}}

\begin{document}

\title{Microscopic optical potentials derived from {\it ab initio} translationally invariant nonlocal one-body densities}

\author{Michael Gennari}
\email{mgennari5216@gmail.com}
\affiliation{University of Waterloo, 200 University Avenue West
Waterloo, Ontario N2L 3G1, Canada\\
TRIUMF, 4004 Wesbrook Mall, Vancouver, British Columbia V6T 2A3, Canada}

\author{Matteo Vorabbi}
\email{mvorabbi@triumf.ca}
\affiliation{TRIUMF, 4004 Wesbrook Mall, Vancouver, British Columbia V6T 2A3, Canada}

\author{Angelo Calci}
\affiliation{TRIUMF, 4004 Wesbrook Mall, Vancouver, British Columbia V6T 2A3, Canada}

\author{Petr Navr\'atil}
\email{navratil@triumf.ca}
\affiliation{TRIUMF, 4004 Wesbrook Mall, Vancouver, British Columbia V6T 2A3, Canada}

\date{\today}

\begin{abstract}

\noindent
{\bf Background:} The nuclear optical potential is a successful tool for the study of nucleon-nucleus elastic scattering and its use has been further extended to
inelastic scattering and other nuclear reactions. The nuclear density of the target nucleus is a fundamental ingredient in the construction of the optical potential and thus plays an important role in the description of the scattering process.

\noindent
{\bf Purpose:} In this work we derive a microscopic optical potential for intermediate energies using {\it ab initio} translationally invariant nonlocal one-body nuclear densities computed within the no-core shell model (NCSM) approach utilizing two- and three-nucleon chiral interactions as the only input.

\noindent
{\bf Methods:} The optical potential is derived at first-order within the spectator expansion of the non-relativistic multiple scattering theory by adopting the impulse approximation. Nonlocal nuclear densities are derived from the NCSM one-body densities calculated in the second quantization. 
The translational invariance is generated by exactly removing the spurious center-of-mass (COM) component from the NCSM eigenstates. 

\noindent
{\bf Results:} The ground state local and nonlocal densities of \textsuperscript{4,6,8}He, \textsuperscript{12}C, and  \textsuperscript{16}O are calculated and applied to optical potential construction. The differential cross sections and the analyzing powers for the elastic proton scattering off these nuclei are then calculated for different values of the incident proton energy. The impact of nonlocality and the COM removal is discussed.

\noindent
{\bf Conclusions:} The use of nonlocal densities has a substantial impact on the differential cross sections and improves agreement with experiment in comparison to results generated with the local densities especially for light nuclei. For the halo nuclei \textsuperscript{6}He and \textsuperscript{8}He, the results for the differential cross section are in a reasonable agreement with the data although a more sophisticated model for the optical potential is required to properly describe the analyzing powers.

\end{abstract}

\pacs{24.10.-i; 24.10.Ht; 24.70.+s; 25.40.Cm;21.60.De;27.10.+h;27.20.+n}

\maketitle

\section{Introduction}

The nuclear optical potential~\cite{hodgson1963} is a successful tool for the investigation of nucleon-nucleus ($NA$) elastic scattering, allowing us to compute the differential
cross section and the spin polarizations in several regions of the nuclear chart and for a wide range of energies. Its use has also been extended to inelastic scattering
calculations and to generate the distorted waves that are used to compute the differential cross section in other nuclear reactions.

Optical potentials can be obtained phenomenologically or microscopically and they are both characterized by a real part describing the nuclear attraction, and an imaginary part
which takes into account the loss of the reaction flux from the elastic channel into the other channels.

Phenomenological potentials assume a certain shape of the nuclear density distribution, which depends on several adjustable parameters that are functions of
the energy and the nuclear mass number~\cite{Varner:1991zz,Koning:2003zz,PhysRevC.80.034608}. These potentials are properly set up in order to optimize the fit to the
experimental data of the $NA$ elastic scattering. Of course, due to the fit, these potentials work very well in situations where experimental data are available, but they lack predictive power.

On the contrary, microscopic optical potentials do not depend on any adjustable parameters making them more appealing for the investigation of new unstable nuclei
where experimental data are not yet available. The computation of such potentials requires, in principle, the solution of the full nuclear many-body problem that has to be solved
using two- and three-nucleon forces as the only input. Unfortunately, such a goal is beyond our actual capabilities and thus some
approximations are needed in order to derive a suitable expression of the optical potential. Several different approaches are currently under development and a complete
list can be found in Ref.~\cite{dickhoff}.

In this paper we adopt the approach based on the nucleon-nucleon ($NN$) $t$ matrix, that was first theoretically justified by Chew \cite{PhysRev.80.196} and
Watson {\it et al.} \cite{PhysRev.89.575,PhysRev.92.291} in the early $50$s, and successively by Kerman, McManus, and Thaler~\cite{Kerman1959551}, who developed the
Watson multiple scattering theory and expressed the optical potential in terms of the free $NN$ scattering amplitudes. Within this framework the optical potential is obtained
from two basic ingredients: the $NN$ $t$ matrix and the matter distribution of the nucleus. Several years later, with the advent of high-accuracy $NN$ potentials, there
was a renewed interest in finding a rigorous treatment of the $NA$ scattering theory and it became possible to compute the $NA$ optical potential directly in momentum space.
Many theoretical efforts~\cite{PhysRevC.41.2257,PhysRevC.44.R1735,PhysRevC.46.279,PhysRevC.48.351,PhysRevC.50.2995,PhysRevLett.63.605,
PhysRevC.41.2188,PhysRevC.42.652,PhysRevC.43.1875,PhysRevC.43.2734,PhysRevC.50.2480,PhysRevC.40.881,PhysRevC.41.814,PhysRevC.44.1569,elster1993,
PhysRevC.47.2242,PhysRevC.48.2956,PhysRevC.51.1418,PhysRevC.52.1992,PhysRevC.56.2080,PhysRevC.57.189,PhysRevC.57.1378} have been devoted to improving
multiple scattering theory in order to obtain a suitable expression for the optical potential capable of reproducing the elastic scattering data.
Similar approaches~\cite{PhysRevC.44.73,dortmans1991,PhysRevC.49.1309,PhysRevC.52.861,dortmans1997,negele2006advances,PhysRevC.52.301,PhysRevC.54.2570,
PhysRevC.66.024602,PhysRevC.76.014613,PhysRevC.78.014608,PhysRevC.84.034606} based on the $g$ matrix were also developed to include the effects
of the nuclear medium on the struck target nucleon interacting with the projectile.

As stated above, the $NA$ optical potential is obtained from the $NN$ $t$ matrix and the nuclear density. Thus, for a consistent calculation it is important to compute
these two quantities using the same $NN$ interaction. In Ref.~\cite{PhysRevC.93.034619} we explored the possibility of using $NN$
chiral potentials~\cite{chiralmachleidt,PhysRevC.88.054002,PhysRevC.87.014322,PhysRevC.91.054311,PhysRevC.75.024311,Machleidt2011,chiralepelbaum}
at the fourth order (N\textsuperscript{3}LO) to compute the optical potential and study the order-by-order convergence in the chiral expansion, comparing the theoretical results for
the scattering observables with the experimental data. In that work we used the factorized model~\cite{PhysRevC.40.881}, with the optical potential computed
as the product of the free $NN$ $t$ matrix and the nuclear density. Within this framework, the calculated potential is nonlocal, but the nonlocality only comes from the
$t$ matrix.

In Ref.~\cite{PhysRevC.96.044001} we performed the same analysis with the same model but using the new $NN$ chiral potentials at the fifth order (N\textsuperscript{4}LO)
developed by Epelbaum, Krebs, and Mei\ss ner~\cite{PhysRevLett.115.122301,Epelbaum2015} and Entem, Machleidt, and 
Nosyk~\cite{PhysRevC.91.014002,PhysRevC.96.024004}, respectively. The conclusion was that convergence in the chiral expansion was basically achieved and
further improvements of the model can only be obtained by reducing the number of approximations made to derive the expression for the optical potential.
Thus, the next step to improve the model consists of following what has been done in the past and computing the optical potential by performing the folding integral of
the $NN$ $t$ matrix with the nuclear density. Of course, the knowledge of a nonlocal density matrix is required.

The purpose of the current work is to use the same $NN$ interaction to compute both the $NN$ $t$ matrix and the nuclear density and to thus achieve another step toward a
consistent microscopic calculation of the $NA$ optical potential. In particular, in this paper we generalize the method introduced in Ref.~\cite{navratil2004translationally} to construct microscopic nonlocal
one-body density within the no-core shell model (NCSM) \cite{barrett2013ab} approach. The NCSM is an {\it ab initio} technique which employs realistic two- and three-nucleon forces and treats all $A$ nucleons in the nucleus as active degrees of freedom. It is particularly well suited for the description of light nuclei, being able to account for many-nucleon correlations producing high-quality wave functions.

The paper is organized as follows: in Sec.~\ref{sec_theoretical_framework}, divided into three subsections, we describe the theoretical framework used to compute the proton-nucleus ($p A$) scattering observables. In Sec.~\ref{sec_optical_potential} we introduce the general scattering problem in
momentum space between the incident proton and the target nucleus, that is described by the many-body Lippmann-Schwinger (LS) equation. This equation is then separated into
one equation for the scattering amplitude and another one for the optical potential, that, after some manipulations and approximations, is reduced to the folding integral
between the free $NN$ $t$ matrix and the nonlocal nuclear density. In Sec.~\ref{sec_ncsm} we then briefly describe the main features of the NCSM and we give the details of the chiral
interactions that we used to perform our calculations. In Sec.~\ref{sec_nonlocaldensity} we give the general expressions for the nonlocal densities and explain how we removed the COM contribution.

In Sec.~\ref{sec_nonlocaldens_results} we present results for the nonlocal and local densities computed with two- and three-body chiral interactions.

In Sec.~\ref{sec_opresults} we discuss differential cross sections and the analyzing powers computed using the nonlocal and local densities presented in Sec.~\ref{sec_nonlocaldens_results}. This section is divided into two sections: in Sec.~\ref{sec_stable_nuclei} we present the scattering observables for
\textsuperscript{4}He, \textsuperscript{12}C, and  \textsuperscript{16}O, for different values of the incident proton energy in the laboratory frame, while in Sec.~\ref{sec_halo_nuclei},
we show the results obtained for the two halo nuclei \textsuperscript{6}He and \textsuperscript{8}He.

Finally, in Sec.~\ref{sec_conclusions} we draw our conclusions.

\section{Theoretical framework}
\label{sec_theoretical_framework}

\subsection{Optical potential}
\label{sec_optical_potential}
The $p A$ elastic scattering problem is described in momentum space by the $(A+1)$-body LS equation for the transition operator
\begin{equation}\label{generallpeq}
T = V + V G_0 (E) T \, ,
\end{equation}
where $V$ represents the external interaction between the incoming proton and the target nucleus and $G_0 (E)$ is the free propagator for the $(A+1)$-nucleon system.
Unfortunately, the solution of Eq.~(\ref{generallpeq}) is beyond our actual capabilities and so some approximations are needed in order to accurately compute the $T$ matrix
without having to treat the full many-nucleon problem. To do this, we follow the standard approach and we introduce the projection operators $P$ and $Q$, where
$P$ projects onto the elastic channel, and they satisfy the completeness relation $P+Q=\id$. With these operators Eq.~(\ref{generallpeq}) can be split in two equations, one for
the $T$ operator,
\begin{equation}\label{firsttamp}
T = U + U G_0 (E) P T \, ,
\end{equation}
and one for the optical potential operator $U$,
\begin{equation}
U = V + V G_0 (E) Q U \, .
\end{equation}
In this work we consider the presence of only two-nucleon forces for the optical potential derivation, and thus the external interaction can be written as
\begin{equation}
V = \sum_{i=1}^A v_{0i} \, ,
\end{equation}
where $v_{0i}$ is the two-body potential describing the interaction between the projectile and the {\it i}th target nucleon. Since we are interested in the elastic scattering process,
we can act with the $P$ operator on both sides of Eq.~(\ref{firsttamp}) to obtain the elastic transition operator,
\begin{equation}\label{elastictop}
T_{\mathrm{el}} \equiv P T P = P U P + P U P G_0 (E) P T_{\mathrm{el}} \, .
\end{equation}
This is a simple one-body equation and can be easily solved with standard techniques once we know the operator $P U P$. Thus, the problem is finding a suitable
expression for the operator $U$ that makes $P U P$ calculable. To do this, we adopt the spectator expansion~\cite{PhysRevC.52.1992} and after introducing the
single-scattering approximation and the impulse approximation, the operator $U$ can be written as
\begin{equation}\label{firstorderop}
U = \sum_{i=1}^A t_{0i} \, ,
\end{equation}
where $t_{0i}$ is the free $NN$ $t$ matrix and satisfies
\begin{equation}
t_{0i} = v_{0i} + v_{0i} g_i t_{0i} \, ,
\end{equation}
with
\begin{equation}
g_i = \frac{1}{E - h_0 - h_i + i \epsilon} \, ,
\end{equation}
being the free $NN$ propagator. Here, $h_0$ is the kinetic energy of the projectile and $h_i$ the kinetic energy of the {\it i}th target nucleon. Acting with the $P$ operator on
Eq.~(\ref{firstorderop}) and after some manipulations, we obtain the expression for the first-order full-folding optical potential
\begin{equation}\label{fullfoldingop}
\begin{split}
U ({\vec q},{\vec K}; \omega ) = &\sum_{\alpha=n,p} \int d^3 {\vec P} \; \eta ({\vec P},{\vec q},{\vec K}) \\
&\times t_{p\alpha} \left[ {\vec q} , \frac{1}{2} \left( \frac{A+1}{A} {\vec K} - {\vec P} \right) ; \omega \right] \\
&\times \rho_{\alpha} \left( {\vec P} - \frac{A-1}{2 A} {\vec q} , {\vec P} + \frac{A-1}{2 A} {\vec q} \right) \, .
\end{split}
\end{equation}
The variables ${\vec q}$ and ${\vec K}$ in Eq.~(\ref{fullfoldingop}) represent the momentum transfer and the total momentum, respectively, while ${\vec P}$
is the integration variable. The quantity $t_{p\alpha}$ is the proton-nucleon free $t$ matrix, $\rho_{\alpha}$ is the nuclear density matrix, and $\eta$
is the M\o ller factor, that imposes the Lorentz invariance of the flux when we pass from the $NA$ to the $NN$ frame in which the $t$ matrices are evaluated.
The energy $\omega$ at which the $t$ matrix is evaluated is fixed at one half the kinetic energy of the incident proton in the laboratory frame.
The structure of the $t$ matrix in the $NN$ frame is given by
\begin{equation}\label{tnnstructure}
t_{p\alpha} ({\vec \kappa}^{\prime} , {\vec \kappa}) = t_{p\alpha}^c ({\vec \kappa}^{\prime} , {\vec \kappa}) + i {\vec \sigma} \cdot \hat{n}_{{\scriptstyle NN}}
t_{p\alpha}^{ls} ({\vec \kappa}^{\prime} , {\vec \kappa}) \, ,
\end{equation}
where ${\vec \kappa}$ and ${\vec \kappa}^{\prime}$ represent the initial and final relative momenta, ${\vec \sigma}$ are the Pauli matrices, and $\hat{n}_{{\scriptstyle NN}}$
is a unit vector defined as $\hat{n}_{{\scriptstyle NN}} = ({\vec \kappa}^{\prime} \times {\vec \kappa})/|{\vec \kappa}^{\prime} \times {\vec \kappa}|$.
Here $t_{p\alpha}^c$ and $t_{p\alpha}^{ls}$ are the central and the spin-orbit parts of the $t$ matrix, respectively. Inserting Eq.~(\ref{tnnstructure}) into Eq.~(\ref{fullfoldingop}) and
after some variable transformations, the general structure of the optical potential is given by
\begin{equation}
U ({\vec q},{\vec K}; \omega )  = U^c ({\vec q},{\vec K}; \omega ) + \frac{i}{2} \vec{\sigma} \cdot \vec{q} \times \vec{K} \, U^{ls} ({\vec q},{\vec K}; \omega )  \, ,
\end{equation}
where $U^c$ and $U^{ls}$ are the central and the spin-orbit parts of the optical potential.
Once Eq.~(\ref{fullfoldingop}) is computed, it is possible to solve Eq.~(\ref{elastictop}) and from its on-shell value we can compute the scattering observables
as specified in Ref.~\cite{PhysRevC.93.034619}. Finally, the Coulomb interaction between the incoming proton and the target nucleus is included as specified
in Refs.~\cite{PhysRevC.44.1569,elster1993}.

\subsection{NCSM}
\label{sec_ncsm}

The evaluation of Eq.~(\ref{fullfoldingop}) requires the knowledge of the nuclear density matrix, that is then folded with the free $NN$ scattering matrix to give the optical potential. The $A$-nucleon eigenstates needed to calculate the one-body density matrix are computed in this work with the {\it ab initio} NCSM method~\cite{barrett2013ab}. Within this framework, nuclei are considered as systems of $A$ nonrelativistic point-like nucleons interacting through realistic inter-nucleon interactions. All nucleons are active degrees of freedom and the translational invariance of observables, as well as the angular momentum and parity of the nucleus under consideration, are conserved. The many-body wave function is expanded over a complete set of antisymmetric $A$-nucleon harmonic oscillator (HO) basis states containing up to $\nmax$ HO excitations above the lowest possible configuration. The basis is further characterized by the frequency $\Omega$ of the HO well.

The NCSM wave functions are computed by diagonalizing the translationally invariant nuclear Hamiltonian, which includes $NN$ and in general also three-nucleon ($3N$) forces:
\begin{equation}\label{NCSM_eq}
\hat{H} \ket{A \lambda J^\pi T} = E_\lambda ^{J^\pi T} \ket{A \lambda J^\pi T} \; ,
\end{equation}
with $\lambda$ labeling eigenstates with identical $J^\pi T$. Convergence of the HO expansion with increasing $N_{\rm max}$ values is accelerated by the use of similarity
renormalization group (SRG) transformations of the $NN$ and $3N$ interactions~\cite{Wegner1994,Bogner2007,PhysRevC.77.064003,Bogner201094,Jurgenson2009evolution}.
While this technique ensures a faster convergence, it may introduce a further dependence on the momentum-decoupling scale $\lambda_{\mathrm{SRG}}$ if the unitarity of the SRG transformation is violated.

In this work, we used the chiral potential at N\textsuperscript{4}LO recently developed by Entem, Machleidt, and Nosyk~\cite{PhysRevC.91.014002,PhysRevC.96.024004} with a cutoff $\Lambda = 500$ MeV employed in the regulator function introduced to deal with the infinities in the LS equation. In the text this will be denoted as NN-N\textsuperscript{4}LO(500). For calculations of nuclear densities, we also included the three-nucleon potential at the next-to next-to leading order (N\textsuperscript{2}LO) with simultaneous local~\cite{Navratil2007} and nonlocal regularization. The three-body component has a local cut-off of $650$ MeV and a nonlocal cut-off of $500$ MeV. The $c_i$ low-energy constants (LECs) were selected as recommended in Ref.~\cite{PhysRevC.96.024004}, while the LECs $c_D$ and $c_E$ were determined in the $A=3,4$ systems. Further details will be given elsewhere. We note that this type of $3N$ interaction was applied for the first time in Ref.~\cite{Leistenschneider:2017mrr} although in combination with a different $NN$ interaction and with different LECs. It will be denoted as 3Nlnl, making the notation for the total interaction utilized for the densities NN-N\textsuperscript{4}LO(500)+3Nlnl.

We have applied the NN-N\textsuperscript{4}LO(500)+3Nlnl to $s$- and $p$-shell nuclei (details will be given elsewhere) and determined that the SRG unitarity is under control for $\lambda_{\mathrm{SRG}}$ in the range of 1.6 to 2.0 fm$^{-1}$ and that the HO frequency of $\hbar\Omega=20$~MeV is suitable for nuclei investigated in this work. To demonstrate convergence of our results, we present basis size ($\nmax$) dependence studies in the subsequent sections. Finally, let us note that the $\nmax=8$ calculations for $^{12}$C and $^{16}$O were obtained using importance-truncated NCSM basis~\cite{Roth2007,Roth2009}.

In addition to nuclear density calculations with SRG-evolved NN-N\textsuperscript{4}LO(500)+3Nlnl interaction, we also computed the $^4$He density with just the bare NN-N\textsuperscript{4}LO(500) to achieve the best consistency of the optical potential construction. See Sec.~\ref{sec_opresults} for further discussion.

\subsection{Nonlocal Nuclear Density}
\label{sec_nonlocaldensity}

With the knowledge of the $A$-nucleon eigenstates it is possible to compute the density matrix needed in Eq.~(\ref{fullfoldingop}). In the current work we generalize the method of Ref.~\cite{navratil2004translationally} to generate nonlocal one-body density matrices and in the following we provide the general formulae. For all the details and the notation we refer the reader to Ref.~\cite{navratil2004translationally}.
We note that the main difference between our approach and the recently presented approach of Ref.~\cite{Burrows2017} resides in the way we remove the COM contribution, which in
this work is done directly in coordinate space, while in Ref.~\cite{Burrows2017} the same is achieved by first transforming the densities to momentum space, removing the COM contributions there and transforming back if needed.

In coordinate representation, the nonlocal form of the nuclear density operator is defined as
\begin{equation}\label{nonlocdens}
\rho_{op}(\vec{r},\vec{r}\,')=\sum_{i=1}^{A} \big( \, \vert \vec{r} \rangle \langle \vec{r}\,' \vert \, \big)^{i}=\sum_{i=1}^{A} \delta(\vec{r}-\vec{r}_i) \delta(\vec{r}\,'-\vec{r}_i') \, .
\end{equation}
The matrix element of this operator between a general initial and final state obtained in the Cartesian coordinate single-particle Slater determinant (SD) basis is written as (compare to Eq. (5) in Ref.~\cite{navratil2004translationally} for the local density)
\begin{equation}\label{eqn:wiCOMnlocdens}
\begin{split}
&{}_{SD}\langle A \lambda_j J_j M_j\, \vert \rho_{op}(\vec{r},\vec{r}\,') \vert \, A \lambda_i J_i M_i \rangle_{SD} \\
& = \sum \frac{1}{\hat{J}_f} ( J_i M_i K k \vert J_f M_f )\,\bigg(Y_{l_1}^*(\hat{r})\,Y_{l_2}^*(\hat{r}\,')\bigg)_k^{(K)} \\
&\quad \times R_{n_1,l_1}\big(\vert \vec{r}\, \vert \big) R_{n_2,l_2}\big(\vert \vec{r}\,'\vert \big) \\
&\quad \times (-1)^{l_1+l_2+K+j_2+\frac{1}{2}} \, \hat{j}_1 \hat{j}_2 \hat{K} \left \{
  \begin{tabular}{ccc}
  $j_2$ & $l_2$ & $\frac{1}{2}$ \\
  $l_1$ & $j_1$ & $K$ \\
  \end{tabular}
\right \} \\
&\quad \times \frac{(-1)}{\hat{K}} {}_{SD} \langle A \lambda_f J_f \vert \vert \,(a_{n_1,l_1,j_1}^{\dagger}\,\tilde{a}_{n_2,l_2,j_2})^{(K)}\,\vert \vert A \lambda_i J_i \rangle_{SD} \,. \\
\end{split}
\end{equation}
In Eq.~(\ref{eqn:wiCOMnlocdens}), the NCSM eigenstates (\ref{NCSM_eq}) have the subscripts $SD$ denoting that we used Slater determinant HO basis that include COM degrees of freedom as opposed to the translationally invariant Jacobi coordinate HO basis~\cite{PhysRevC.61.044001}. The isospin and parity quantum numbers are supressed for simplicty. Further, $\hat{\eta}=\sqrt{2\eta+1}$ and 
$R_{n,l}(\vert \vec{r} \vert)$ is the radial HO wave function with the oscillator length parameter $b=\sqrt{\frac{\hbar}{m\Omega}}$, where $m$ is the nucleon mass. The one-body density matrix elements are introduced in the second-quantization,
${}_{SD} \langle A \lambda_f J_f \vert \vert \,(a_{n_1,l_1,j_1}^{\dagger}\,\tilde{a}_{n_2,l_2,j_2})^{(K)}\,\vert \vert A \lambda_i J_i \rangle_{SD}$.
Both $\vec{r}$ and $\vec{r}\,'$ are measured from the center of the HO potential well. Consequently, the density contains a spurious COM component.

We require the removal of the COM component from the nonlocal density. This is enabled by the factorization of the Slater determinant and Jacobi eigenstates,
\begin{equation}\label{eq:sd_jac_fac}
\begin{split}
&\langle \vec{r}_1 \dots \vec{r}_A \vec{\sigma}_1 \dots \vec{\sigma}_A \vec{\tau}_1 \dots \vec{\tau}_A \vert A \lambda J M \rangle_{SD}= \\
&\qquad \langle \vec{\xi}_1 \dots \vec{\xi}_{A-1} \vec{\sigma}_1 \dots \vec{\sigma}_A \vec{\tau}_1 \dots \vec{\tau}_A \vert A \lambda JM \rangle \phi_{000}(\vec{\xi}_0) \, , \\
\end{split}
\end{equation}
with COM component, labeled in Eq.~(\ref{eq:sd_jac_fac}) as $\phi_{000}(\vec{\xi}_0)$, given as the $N=0$ HO state with $\vec{\xi}_0$ proportional to the $A$-nucleon COM coordinate. The translational invariance can be then obtained by employing the same procedure outlined for local densities
in Ref.~\cite{navratil2004translationally} on the COM contaminated nonlocal density. The matrix element of the translationally invariant operator, $\rho_{op}^{trinv}(\vec{r}-\vec{R},\vec{r}^{\, \prime}-\vec{R})$, between general initial and final states is then given by (compare to Eq. (16) in Ref.~\cite{navratil2004translationally} for the local density)
\begin{equation}\label{eqn:trinvnlocdens}
\begin{split}
&\langle A \lambda_j J_j M_j\, \vert \rho_{op}^{trinv}(\vec{r}-\vec{R},\vec{r}\,'-\vec{R}) \vert \, A \lambda_i J_i M_i \rangle \\
& = \Big(\frac{A}{A-1}\Big)^{\frac{3}{2}}\sum \frac{1}{\hat{J}_f} ( J_i M_i K k \vert J_f M_f ) \\
&\quad \times \big(M^K\big)_{nln'l',n_1l_1n_2l_2}^{-1} \, \bigg(Y_l^*(\widehat{\vec{r}-\vec{R}})\,Y_{l'}^*(\widehat{\vec{r}\,'-\vec{R}})\bigg)_k^{(K)} \\
&\quad \times R_{n,l}\Big(\sqrt{\frac{A}{A-1}} \vert \vec{r}-\vec{R} \vert \Big) R_{n',l'}\Big(\sqrt{\frac{A}{A-1}} \vert \vec{r}\,'-\vec{R} \vert \Big) \\ 
&\quad \times (-1)^{l_1+l_2+K+j_2-\frac{1}{2}} \, \hat{j}_1 \hat{j}_2 \left \{
  \begin{tabular}{ccc}
  $j_1$ & $j_2$ & $K$ \\
  $l_2$ & $l_1$ & $\frac{1}{2}$ \\
  \end{tabular}
\right \} \\
&\quad \times {}_{SD} \langle A \lambda_f J_f \vert \vert \,(a_{n_1,l_1,j_1}^{\dagger}\,\tilde{a}_{n_2,l_2,j_2})^{(K)}\,\vert \vert A \lambda_i J_i \rangle_{SD} \\
\end{split}
\end{equation}
where
\begin{equation}
\label{Mk}
\begin{split}
&\big( M^K \big)_{nln'l',n_1l_1n_2l_2} \\
&= \sum_{N_1,L_1} (-1)^{l+l'+K+L_1} \left \{
  \begin{tabular}{ccc}
  $l_1$ & $L_1$ & $l$ \\
  $l'$ & $K$ & $l_2$ \\
  \end{tabular}
\right \} \hat{l} \hat{l'} \\
&\quad \times \langle nl00l \vert N_1 L_1 n_1 l_1 l \rangle_{\frac{1}{A-1}} \langle n'l'00l' \vert N_1 L_1 n_2 l_2 l' \rangle_{\frac{1}{A-1}} \, \, .
\end{split}
\end{equation}
In Eq.~(\ref{eqn:trinvnlocdens}), the $R_{n,l}\Big(\sqrt{\frac{A}{A-1}} \vert \vec{r}-\vec{R} \vert \Big)$ is the radial harmonic oscillator wave function in terms of a relative
Jacobi coordinate, $\vec{\xi}=-\sqrt{\frac{A}{A-1}}(\vec{r}-\vec{R})$. The $\big( M^K \big)_{nln'l',n_1l_1n_2l_2}$ matrix (\ref{Mk}) introduced in Ref.~\cite{navratil2004translationally} includes generalized harmonic oscillator brackets of the form $\langle nl00l \vert N_1 L_1 n_1 l_1 l \rangle_{d}$ corresponding to a two particle system with a mass ratio of $d$, as outlined in Ref.~\cite{trlifaj1972simple}.

The nonlocal density expressions presented here can be related to the local density ones in Ref.~\cite{navratil2004translationally} by setting
\begin{equation}
\rho(\vec{r}) = \rho(\vec{r},\vec{r})  \, .
\end{equation}
For both the COM contaminated and translationally invariant nonlocal density we recover the corresponding local density, as expected. This procedure is detailed in the Appendix for the case of the translationally invariant density. The normalization of the nonlocal density is consistent with Ref.~\cite{navratil2004translationally} such that the integral of the local form,
\begin{equation}\label{eq:dens_norm}
\int d\vec{r} \, \langle A \lambda J M \vert \rho_{op}(\vec{r}, \vec{r}) \vert A \lambda J M \rangle = A \,,
\end{equation}
returns the number of nucleons for both (\ref{eqn:wiCOMnlocdens}) and (\ref{eqn:trinvnlocdens}).

Finally, let us note that the proton and neutron densities are obtained simply by introducing ($\frac{1}{2}\pm t_{zi}$) factors, respectively, in Eq. (\ref{nonlocdens}), which then results in the creation and anihilation operators aquiring a proton or neutron index as the COM operators commute with isospin operators. The normalization (\ref{eq:dens_norm}) then changes to $Z$ and $N$ for the proton and neutron density respectively. 

\subsection{Nonlocal density in momentum space}
\label{sec_densmomspace}

In Sec.~\ref{sec_nonlocaldensity} we presented the general expressions for the nonlocal densities in coordinate space, but the evaluation of Eq.~(\ref{fullfoldingop}) for the
optical potential requires the knowledge of the ground-state density in momentum space. In the following we show how this was done.
For the ground state of even-even nuclei, considered in this work, the angular momenta $J_i$ and $J_f$ in Eq.~(\ref{eqn:wiCOMnlocdens}) and Eq.~(\ref{eqn:trinvnlocdens})
are equal to zero: this gives $k = K = 0$ and consequently $l^{\prime} = l$. Thus, Eq.~(\ref{eqn:wiCOMnlocdens}) and Eq.~(\ref{eqn:trinvnlocdens}) can be expressed
in a general form as
\begin{equation}
\rho (\vec{r},\vec{r}^{\, \prime}) = \sum_l \rho_l (r,r^{\prime}) \bigg(Y_l^*(\hat{r}) \,Y_l^* (\hat{r}^{\prime}) \bigg)_0^{(0)} \, ,
\end{equation}
where $\rho_l (r,r^{\prime})$ is obtained summing the radial part over all the other quantum numbers. The angular part can be easily evaluated as
\begin{equation}
\bigg(Y_l^*(\hat{r}) \,Y_l^* (\hat{r}^{\prime}) \bigg)_0^{(0)} = {(-1)}^l \frac{\sqrt{2 l +1}}{4 \pi} P_l (\cos \omega) \, ,
\end{equation}
where $P_l$ are the Legendre polynomials and $\omega$ is the angle between $\vec{r}$ and $\vec{r}^{\, \prime}$. In momentum space, the expression of the density is given by
\begin{equation}
\rho (\vec{p},\vec{p}^{\, \prime}) = \frac{1}{2 \pi^2} \sum_l \rho_l (p,p^{\prime}) {(-1)}^l \sqrt{2 l +1} P_l (\cos \gamma) \, ,
\end{equation}
where $\gamma$ is the angle between $\vec{p}$ and $\vec{p}^{\, \prime}$. The radial part $\rho_l (p,p^{\prime})$ is finally obtained as
\begin{equation}
\rho_l (p,p^{\prime}) = \int_0^{\infty} d r r^2 \int_0^{\infty} d r^{\prime} r^{\prime \, 2} j_l (p r) \rho_l (r,r^{\prime}) j_l (p^{\prime} r^{\prime}) \, ,
\end{equation}
where $j_l$ are the spherical Bessel functions.

\section{Nonlocal Density Results}
\label{sec_nonlocaldens_results}

In this section we show the results for the nonlocal densities obtained from the NCSM wave functions and using the approach described in Sec.~\ref{sec_nonlocaldensity}. The SRG-evolved NN-N4LO(500)+3Nlnl interaction was used in all results discussed in the section.
As a test of the importance of COM removal, we computed for \textsuperscript{4,6,8}He,  \textsuperscript{12}C, and \textsuperscript{16}O the translational invariant and
COM contaminated nuclear densities given by Eq.~(\ref{eqn:trinvnlocdens}) and Eq.~(\ref{eqn:wiCOMnlocdens}), respectively. Figure plots of the COM contaminated density are labeled \textit{wiCOM} while the translationally
invariant density plots are labeled \textit{trinv}. The ground-state densities of the nuclei are shown with all angular dependence factorized out for plotting.

\begin{figure}[t]
\includegraphics[width=0.5\textwidth]{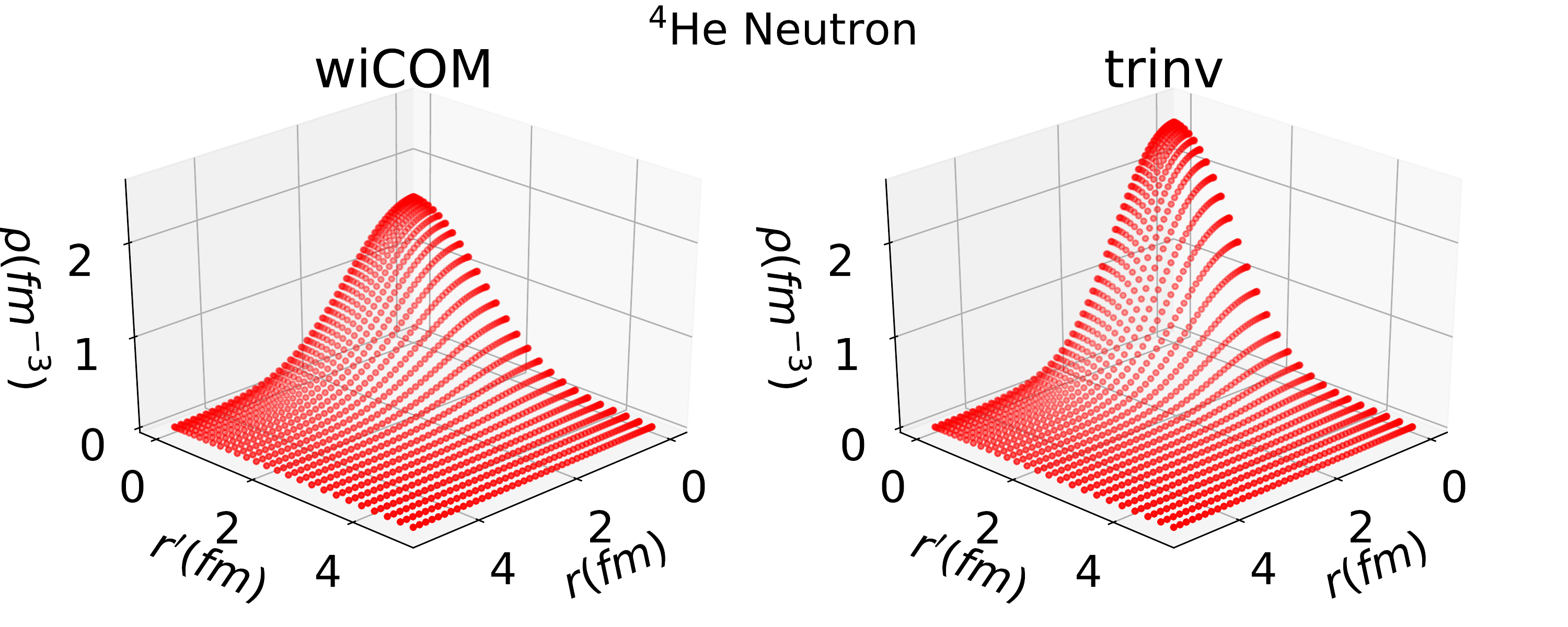}
\caption{\label{fig:nonlocal_nHe4} Ground-state \textsuperscript{4}He nonlocal neutron density calculated with an $\nmax = 14$ basis space, an oscillator frequency
of $\hb = 20$ MeV, and a flow parameter of $\lambda_{\mathrm{SRG}}=2.0$ fm\textsuperscript{-1}.}
\end{figure}

To appreciate the significance of spurious COM removal in light nuclei, consider the comparison between the {\it wiCOM} and {\it trinv} nonlocal density of
\textsuperscript{4}He shown in Fig. \ref{fig:nonlocal_nHe4}. An $\nmax = 14$ basis space is used with a flow parameter $\lambda_{\mathrm{SRG}} = 2.0$ fm\textsuperscript{-1}.
The tremendous difference between the {\it trinv} density and the {\it wiCOM} density is easily recognizable at small $r$ and $r^{\prime}$. We notice that the
{\it trinv} density has sharper features at peaks and tends to decay more rapidly than the {\it wiCOM} density. The COM contamination appears to suppress
the nuclear density at small $r$ and $r^{\prime}$ values.

\begin{figure}[t]
\includegraphics[width=0.5\textwidth]{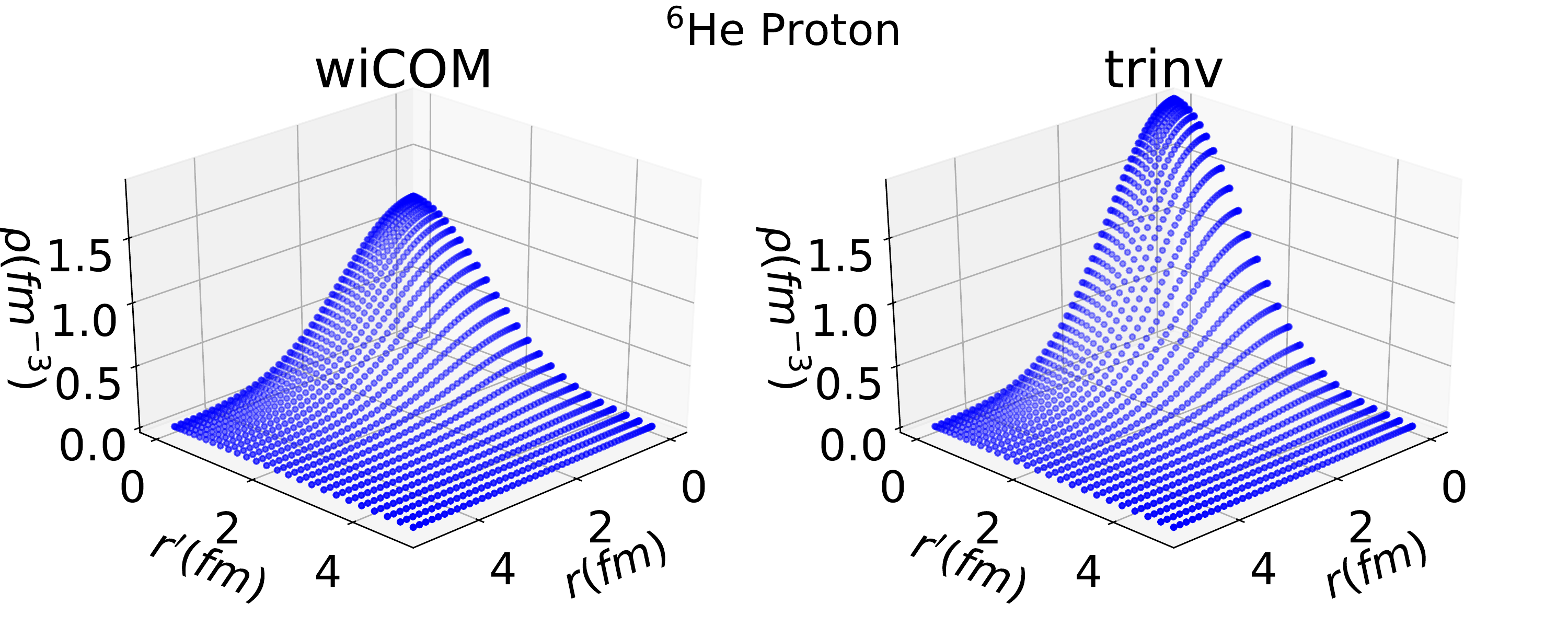}
\includegraphics[width=0.5\textwidth]{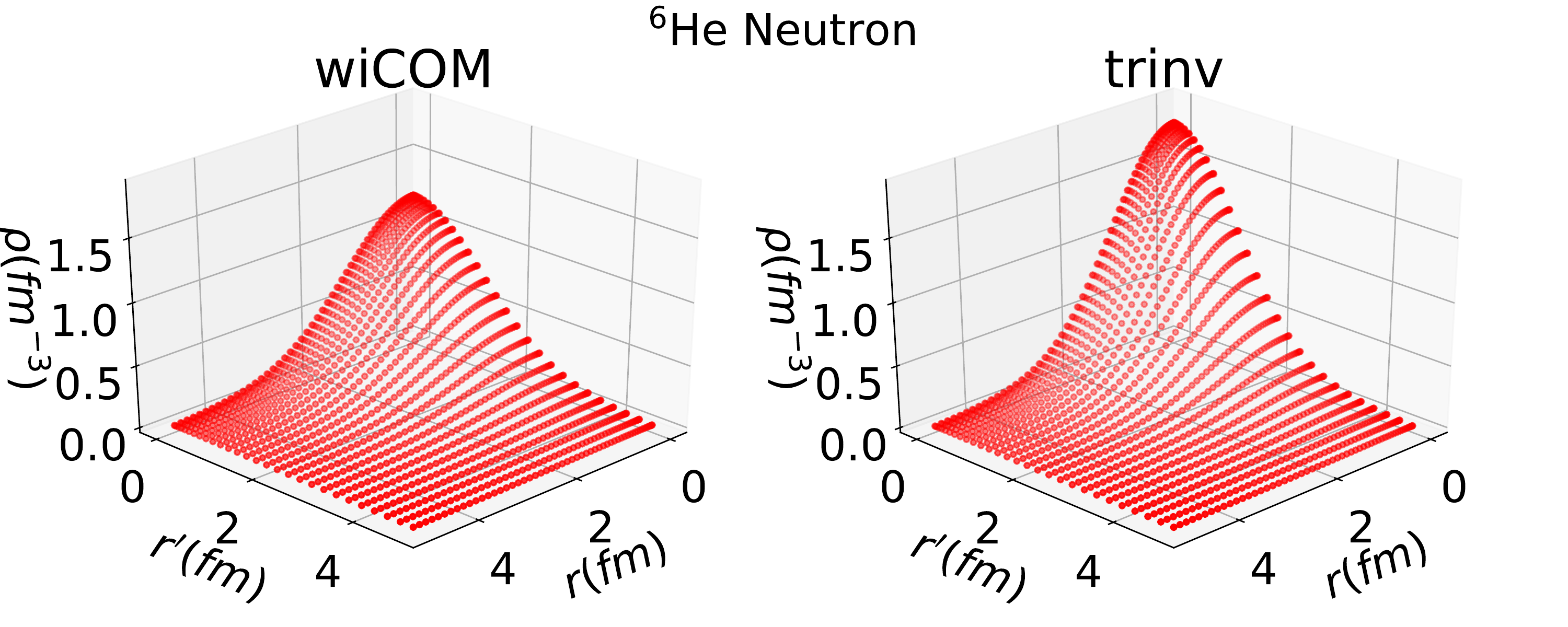}
\caption{\label{fig:nonlocalHe6} Ground-state \textsuperscript{6}He proton and neutron nonlocal densities calculated with a $\nmax = 12$ basis space, an oscillator frequency
of $\hb = 20$ MeV, and a flow parameter of $\lambda_{\mathrm{SRG}} = 2.0$ fm\textsuperscript{-1}. Proton densities are shown in blue and neutron densities are shown in red.}
\end{figure}

\begin{figure}[t]
\includegraphics[width=0.5\textwidth]{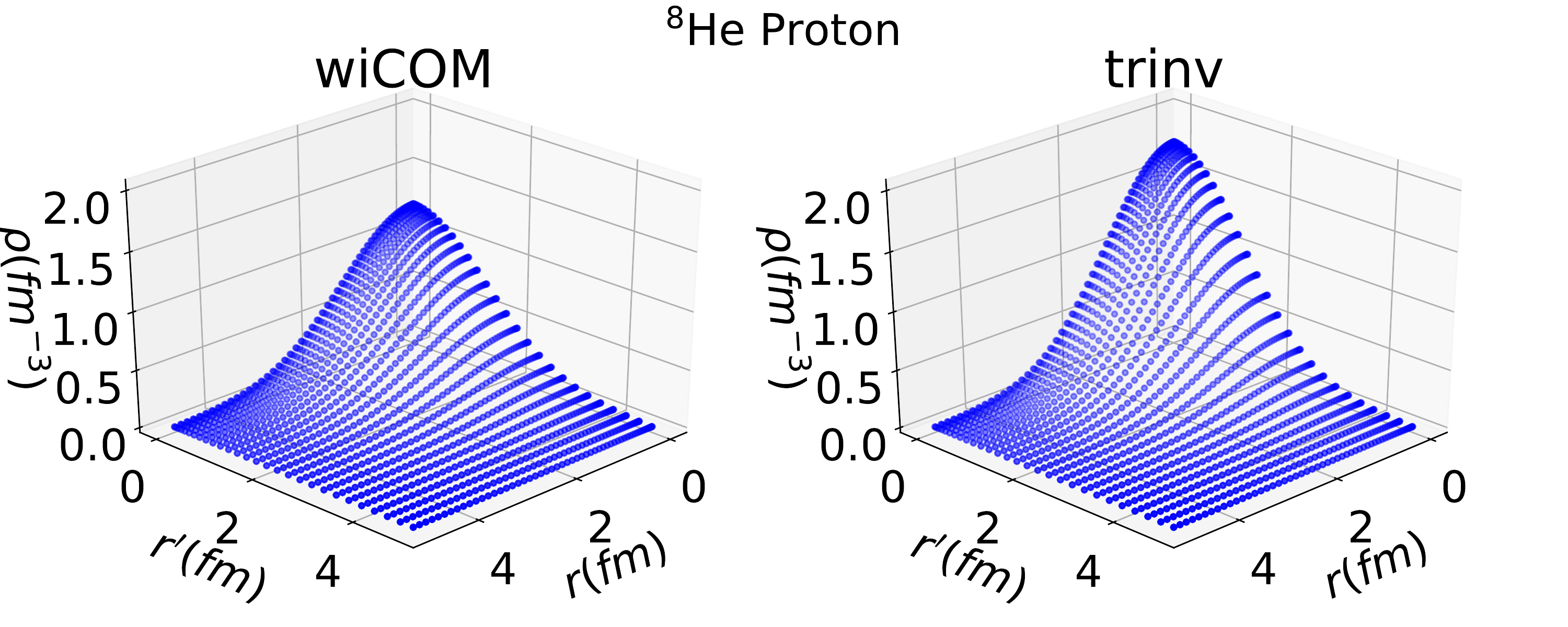}
\includegraphics[width=0.5\textwidth]{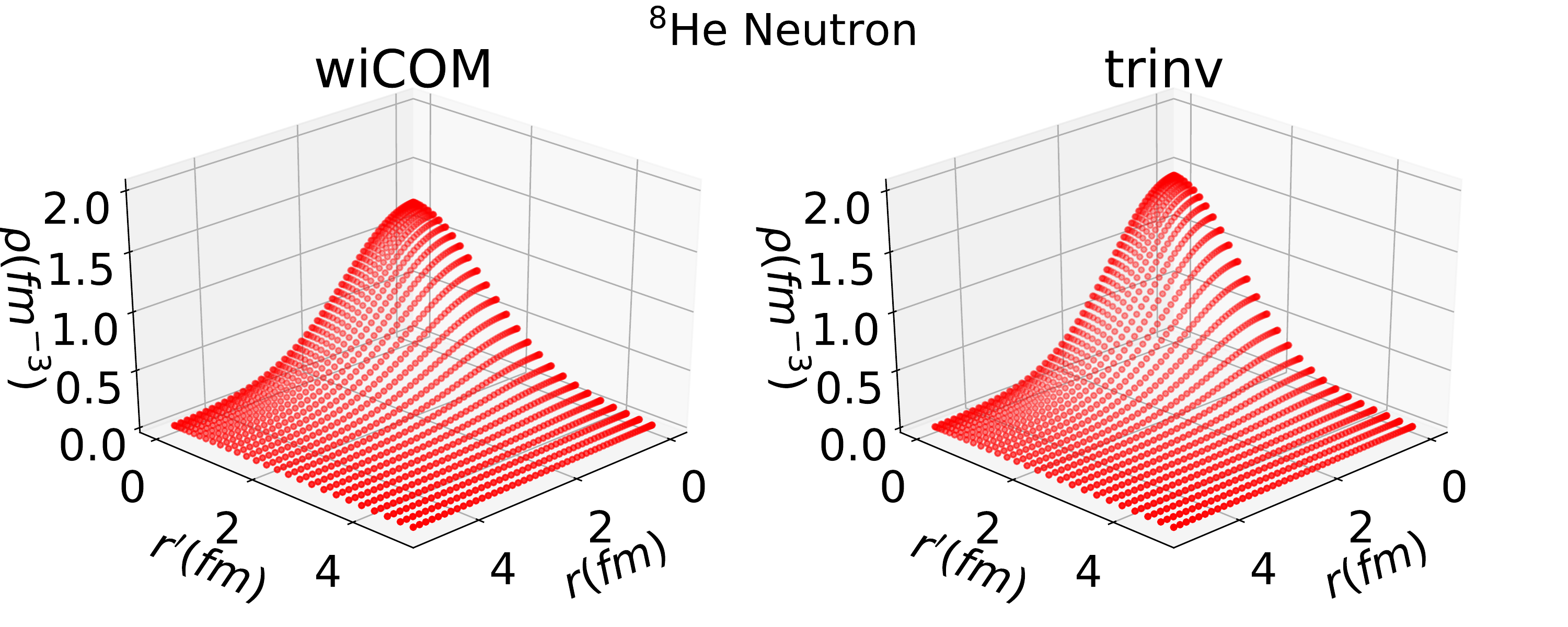}
\caption{\label{fig:nonlocalHe8} Ground-state \textsuperscript{8}He proton and neutron nonlocal densities calculated with a $\nmax = 10$ basis space, an oscillator frequency
of $\hb = 20$ MeV, and a flow parameter of $\lambda_{\mathrm{SRG}} = 2.0$ fm\textsuperscript{-1}. Proton densities are shown in blue and neutron densities are shown in red.}
\end{figure}

In Fig. \ref{fig:nonlocalHe6} we present the proton and neutron nonlocal densities for \textsuperscript{6}He using a $\nmax = 12$ basis space with a flow parameter
$\lambda_{\mathrm{SRG}}=2.0$ fm\textsuperscript{-1}. As in the case of \textsuperscript{4}He, the translationally invariant density behaves significantly different from the spurious
COM contaminated density. We still see that the COM tends to smooth the density over larger $r$ and $r^{\prime}$ values, suppressing it for small $r$ and $r^{\prime}$.
However, we see a minor reduction in peak amplitude and sharpness when compared to the differences observed in \textsuperscript{4}He. Notably, the COM term diminishes with $A$ so we expect a reduction in the importance of its removal as we go to higher $A$-nucleon systems. This trend is further noticeable in
Fig.~\ref{fig:nonlocalHe8} which shows results for the nonlocal density of \textsuperscript{8}He using the same $\lambda_{\mathrm{SRG}}$ parameter and a $\nmax = 10$ basis space.

\begin{figure}[t]
\includegraphics[width=0.5\textwidth]{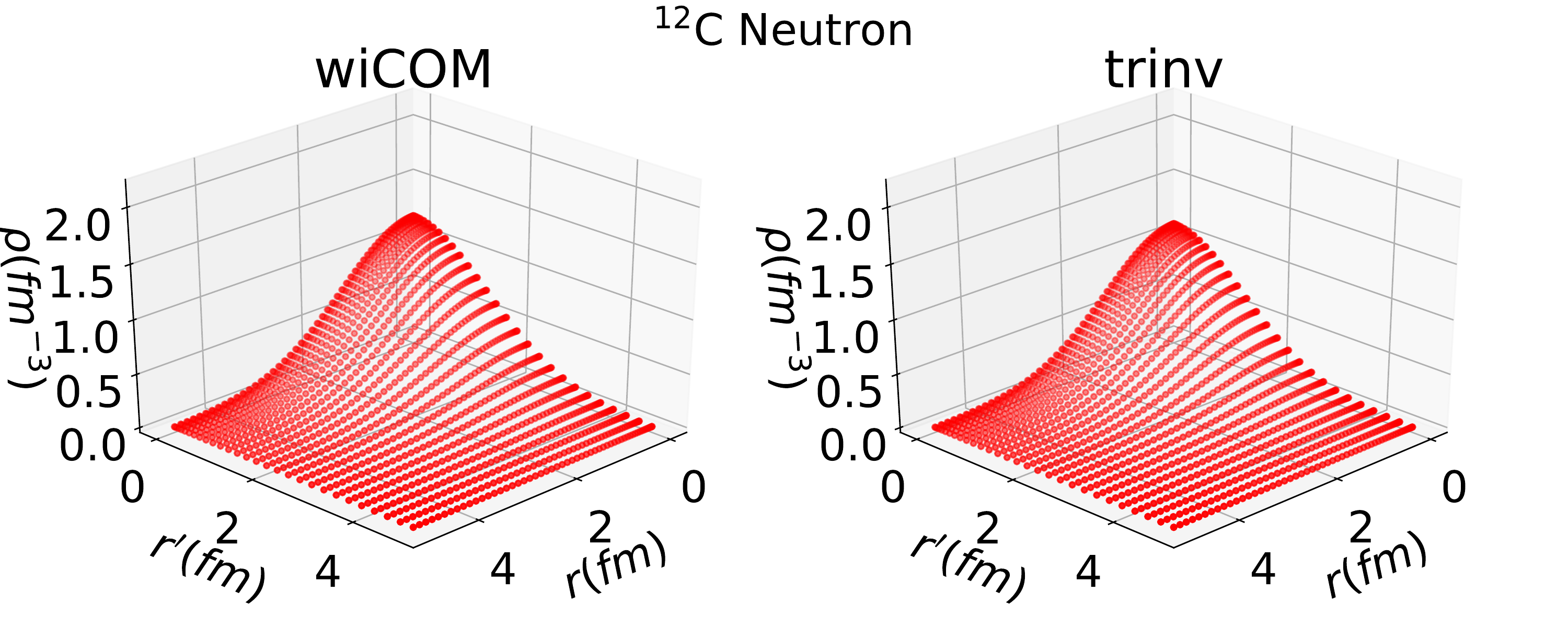}
\includegraphics[width=0.5\textwidth]{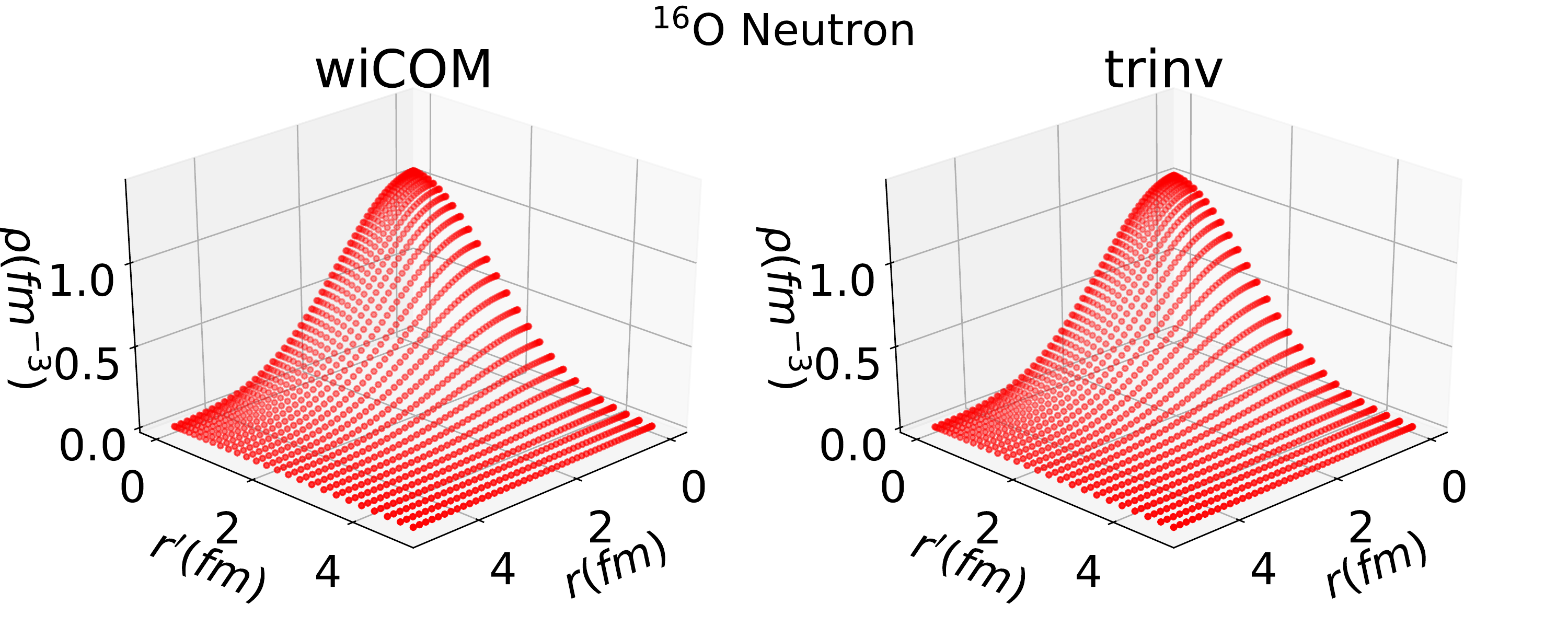}
\caption{\label{fig:nonlocal_nO16} Ground-state nonlocal neutron densities calculated with an $\nmax = 8$ importance truncated NCSM basis for \textsuperscript{12}C and \textsuperscript{16}O. An oscillator frequency of $\hb = 20$ MeV, and a flow parameter of $\lambda_{\mathrm{SRG}} =1.8$ fm\textsuperscript{-1}
and $\lambda_{\mathrm{SRG}} = 2.0$ fm\textsuperscript{-1} are used, respectively.}
\end{figure}

In Fig. \ref{fig:nonlocal_nO16} we present the nonlocal neutron density for \textsuperscript{12}C and \textsuperscript{16}O. For both nuclei we use $\nmax=8$ basis space with $\lambda_{\mathrm{SRG}}=1.8$ fm\textsuperscript{-1} and 2.0 fm\textsuperscript{-1} for \textsuperscript{12}C and \textsuperscript{16}O, respectively. Unlike the results for \textsuperscript{4}He and \textsuperscript{6}He, we see only minor effects from the process of COM removal. As expected, with increasing $A$-nucleon number the effect of COM removal is further suppressed to a point at
which it becomes difficult to differentiate between the COM contaminated density and the translationally invariant density.

\begin{figure}[t]
\includegraphics[width=0.52\textwidth]{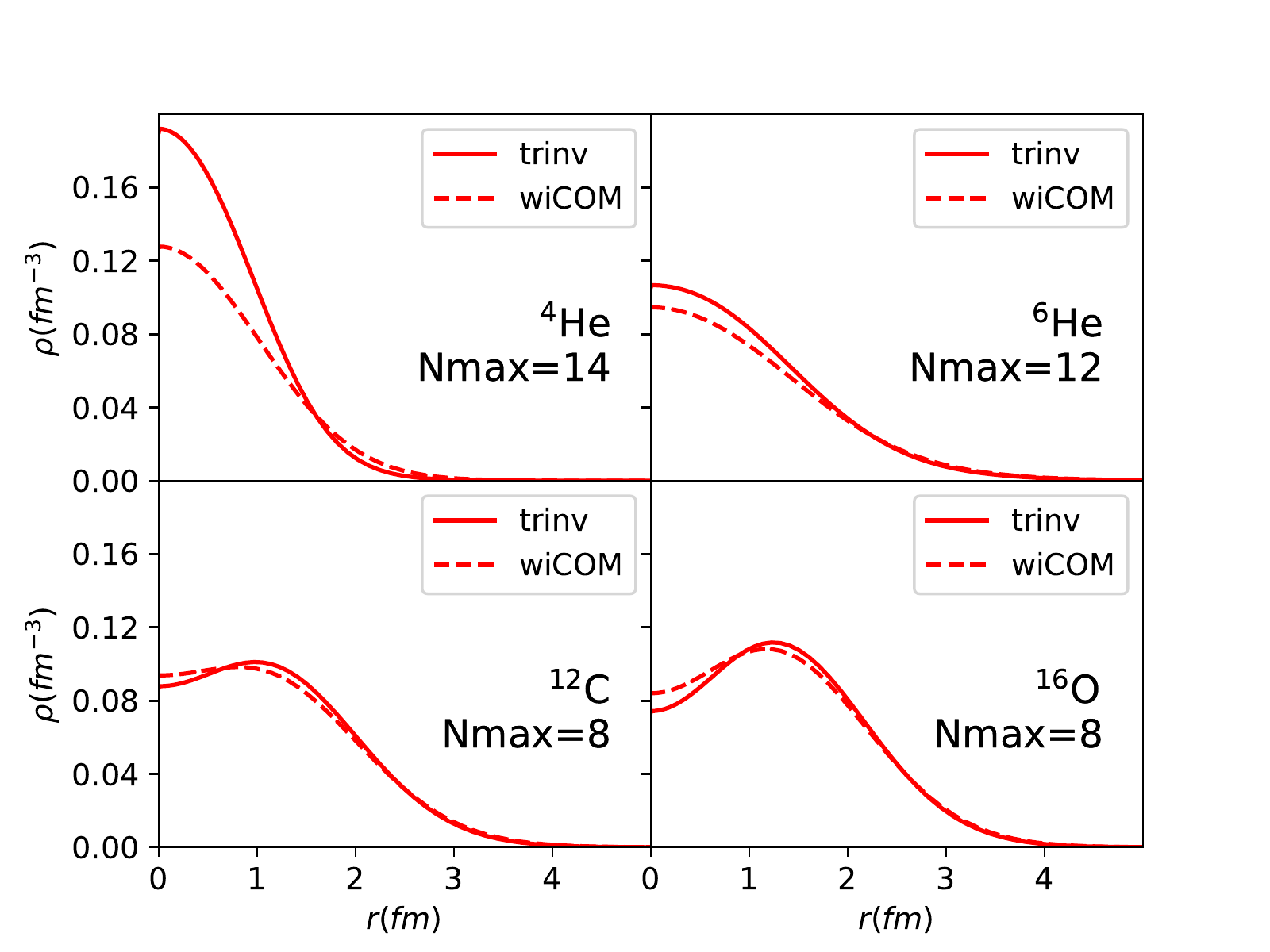}
\caption{\label{fig:local_comp} Comparison between COM contaminated (dashed {\it wiCOM}) and translationally invariant (solid {\it trinv}) local neutron densities for \textsuperscript{4}He,
\textsuperscript{6}He, \textsuperscript{12}C, and \textsuperscript{16}O.}
\end{figure}

However, if we investigate the local densities of these nuclei in Fig. \ref{fig:local_comp}, we see more apparent differences between the COM contaminated and translationally
invariant densities in higher $A$-nucleon systems.

Given the sizeable differences between the translationally invariant and COM contaminated density in the case of light nuclei, we expect a significant impact on observables
related to the nonlocal density. This will be further discussed in Sec.~\ref{sec_opresults}. It is expected that some observables may amplify the effects observed and thus better
gauge the importance of COM removal being performed on nuclei such as \textsuperscript{16}O. The nonlocal densities of these heavier nuclei must be further investigated in order to fully determine the importance of COM removal.

\begin{figure}[t]
\includegraphics[width=0.52\textwidth]{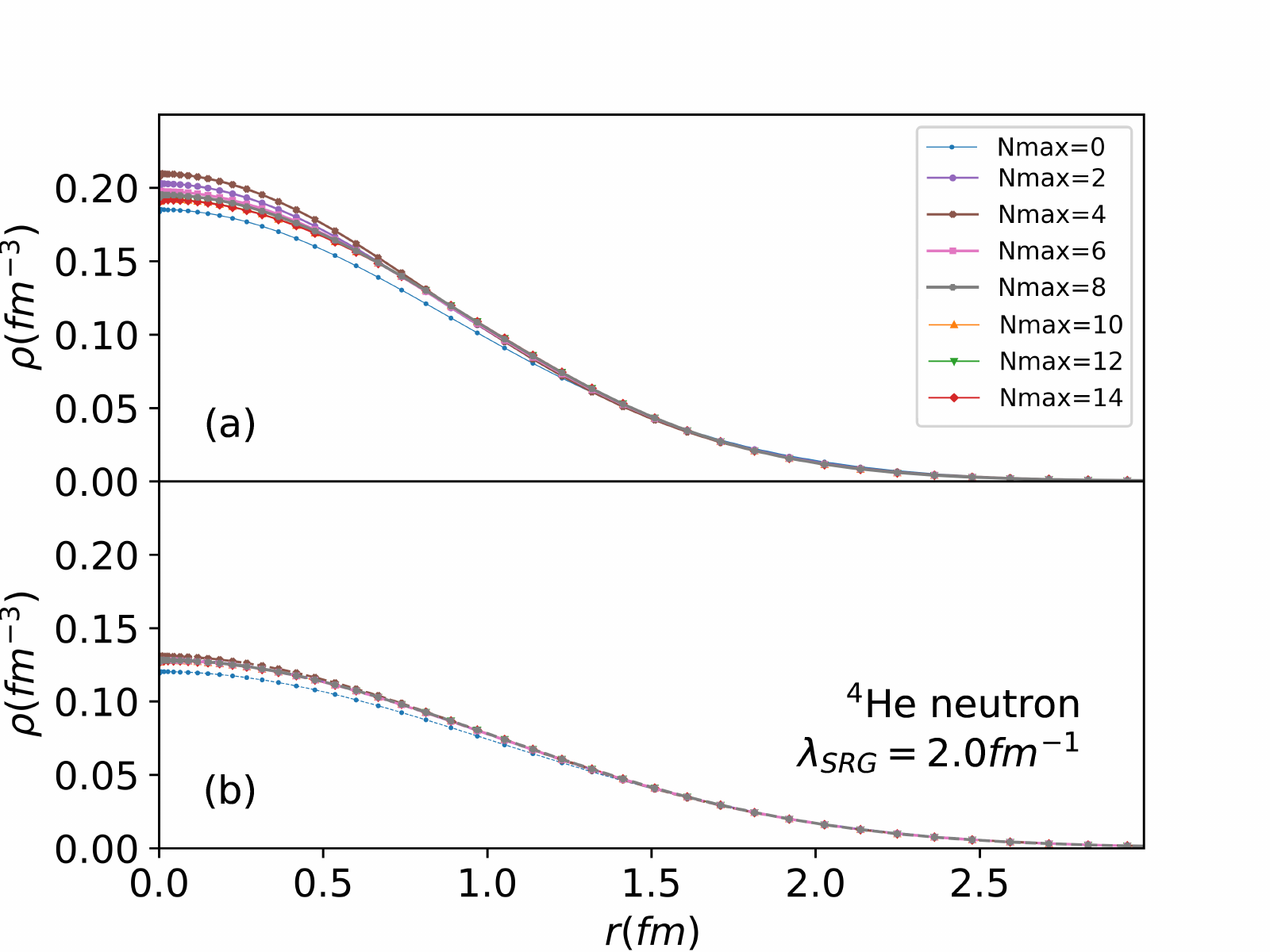}
\caption{\label{fig:convergence_He4} $\nmax$ convergence for the translationally invariant (a) and the COM contaminated (b)
local neutron densities for \textsuperscript{4}He.}
\end{figure}

\begin{figure}[t]
\includegraphics[width=0.49\textwidth]{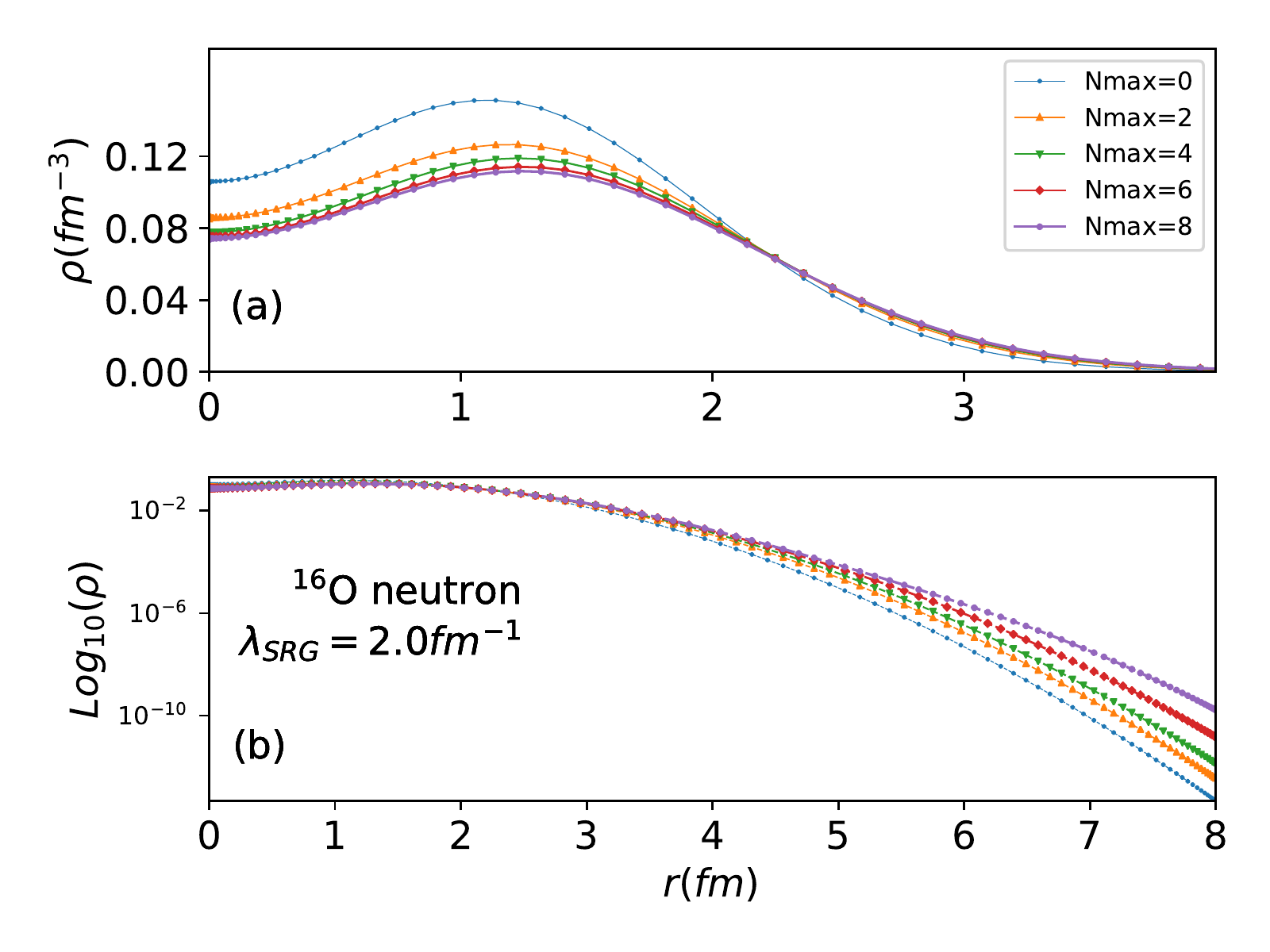}
\caption{\label{fig:convergence_O16} {\bf Panel (a):} standard $\nmax$ convergence comparison. {\bf Panel (b):} long-range $\nmax$ comparison.
Both plots use the translationally invariant neutron density of \textsuperscript{16}O.}
\end{figure}

In Fig.~\ref{fig:convergence_He4} and Fig.~\ref{fig:convergence_O16} we present $\nmax$ convergence plots for \textsuperscript{4}He and \textsuperscript{16}O.
We see that for \textsuperscript{4}He we achieve rapid convergence with the NN-N4LO(500)+3Nlnl interaction and a basis size of $\nmax = 10$.
Even in \textsuperscript{16}O we see good convergence trends in the local density at the basis size of $\nmax = 8$ in the whole range up to $\sim 6$~fm.

Since we use the harmonic oscillator basis, all of our densities have ``unphysical'' asymptotic behaviour due to the Gaussian tail resulting from the basis expansion.
Naturally, the tail should behave exponentially. In the logarithmic plot in panel (b) of Fig.~\ref{fig:convergence_O16}, we can see a very slow convergence at long distances. 
Notice though, that these effects occur maximally on the order of $\sim 10^{-8}$ and so are not of serious concern in particular in the present application to high-energy nucleon-nucleus scattering.

\section{Scattering observables}
\label{sec_opresults}

In this section we present the results for the scattering observables computed with Eq.~(\ref{fullfoldingop}) and using the density matrices shown in the previous section.
The density matrices are related to the density profiles $\rho_{\alpha} (q)$ of the nucleus by
\begin{equation}\label{density_profile}
\rho_{\alpha} (q) = \int d^3 {\vec P} \; \rho_{\alpha} \left( {\vec P} - \frac{A-1}{2 A} {\vec q} , {\vec P} + \frac{A-1}{2 A} {\vec q} \right) \, ,
\end{equation}
and they respect the normalization of Eq.~(\ref{eq:dens_norm}), that in momentum space it is given by $\rho_{\alpha} (q=0) = N,Z$.
The {\it trinv} and {\it wiCOM} density matrices, that are used in Eq.~(\ref{fullfoldingop}) and in Eq.~(\ref{density_profile}), are obtained as specified in Sec.~\ref{sec_densmomspace}.
Moreover, in order to assess how these new nonlocal densities improve the calculation of the optical potential, we compare the results computed with Eq.~(\ref{fullfoldingop})
with those obtained using the simpler factorized optical potential
\begin{equation}\label{factorizedop}
U ({\vec q},{\vec K}; \omega ) = \eta ({\vec q},{\vec K}) \sum_{\alpha = n,p} t_{p\alpha} \left[{\vec q},\frac{A+1}{2A} {\vec K} ; \omega \right] \, \rho_{\alpha} (q) \, ,
\end{equation}
where $\rho_{\alpha} (q)$ is the same of Eq.~(\ref{density_profile}), but in this case it is obtained computing the Fourier transform of the local density~\cite{navratil2004translationally}
in coordinate space:
\begin{equation}\label{ftlocaldens}
\rho_{\alpha} (q) = 4 \pi \int_0^{\infty} d r r^2 j_0 (q r) \rho_{\alpha} (r) \, .
\end{equation}
Of course, it is possible to compute $\rho_{\alpha} (q)$ using Eq.~(\ref{density_profile}) and then calculate the scattering observables using
Eq.~(\ref{factorizedop}). In this case, the only difference between the results computed with Eq.~(\ref{fullfoldingop}) and Eq.~(\ref{factorizedop}) comes from the treatment
of the folding integral and not from the densities. Below we provide an example for such a calculation.

Finally, it is important to notice that Eq.~(\ref{fullfoldingop}) and Eq.~(\ref{factorizedop}) are both obtained including only the $NN$ interaction and they do not contain any
contribution from the three-nucleon forces. This makes our results not fully consistent, but, at present, the inclusion of the three-nucleon forces in the model for the optical
potential represents a very difficult task. Thus, our results have been obtained using either Eq.~(\ref{fullfoldingop}) or Eq.~(\ref{factorizedop}), where the scattering part of the optical
potential was computed with the only NN-N\textsuperscript{4}LO(500) interaction, while the nuclear densities were computed with the NN-N\textsuperscript{4}LO(500)+3Nlnl interaction.
Another important aspect is represented by the SRG method, that has been used to evolve the $NN$ interaction for the calculation of the densities, but not for the calculation
of the $NN$ $t$ matrix, where the bare interaction has been employed. Below we discuss this consistency problem with more details.

\subsection{Results for \textsuperscript{4}He, \textsuperscript{12}C, and \textsuperscript{16}O}
\label{sec_stable_nuclei}

\begin{figure}[t]
\begin{center}
\includegraphics[scale=0.32]{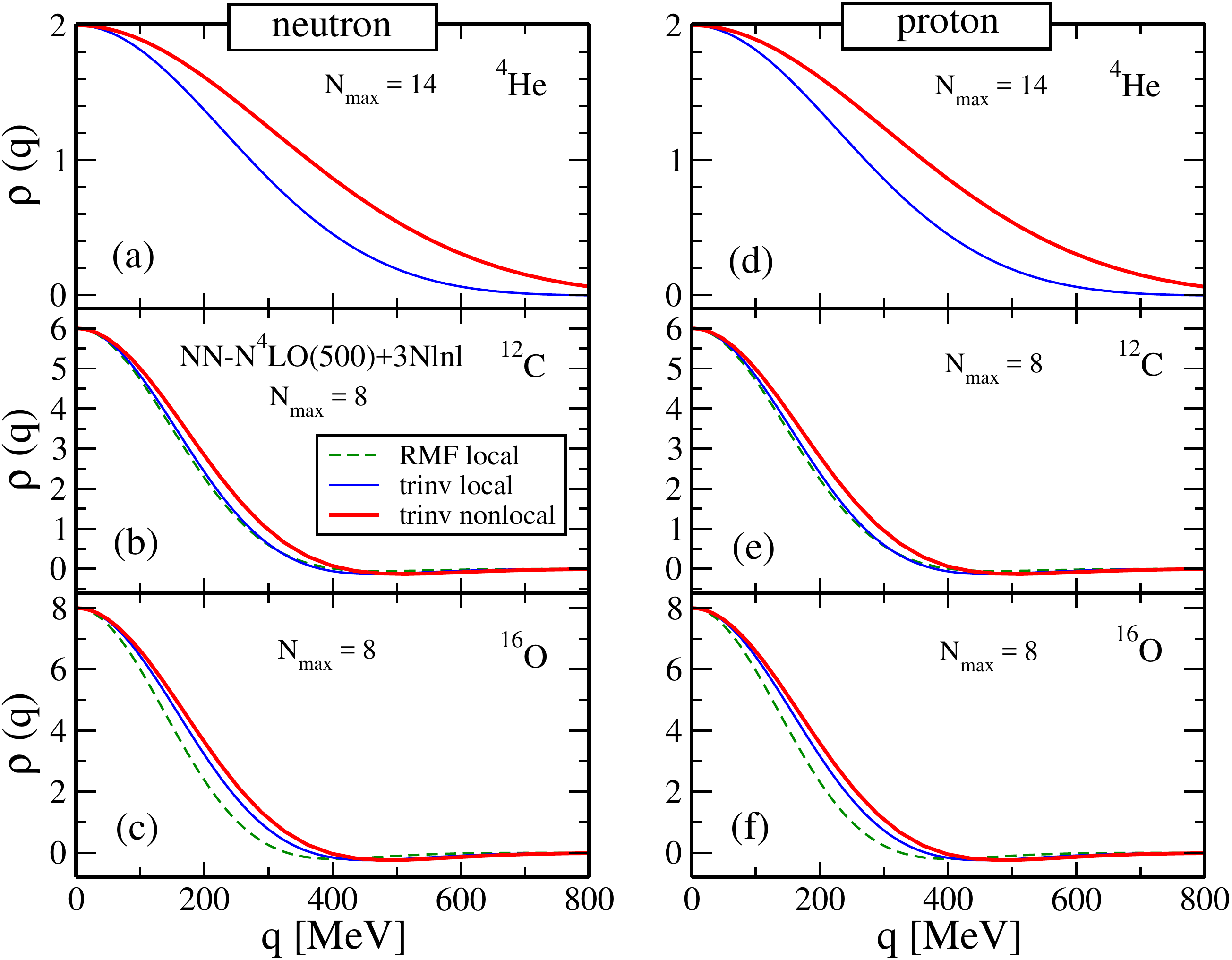}
\caption{\label{density_profiles} (Color online) Neutron and proton nuclear density profiles for \textsuperscript{4}He, \textsuperscript{12}C, and \textsuperscript{16}O. The results
named as "local" were obtained from Eq.~(\ref{ftlocaldens}), while the results named as "nonlocal" were obtained from Eq.~(\ref{density_profile}).
For \textsuperscript{12}C and \textsuperscript{16}O we also include the nuclear profile computed within a Relativistic Mean-Field description of spherical
nuclei. The densities were computed with the NN-N\textsuperscript{4}LO(500)+3Nlnl interaction with $\hb = 20$ MeV, $\lambda_{\mathrm{SRG}} = 2.0$ fm\textsuperscript{-1}, and
for different values of $\nmax$.}
\end{center}
\end{figure}

\begin{figure}[t]
\begin{center}
\includegraphics[scale=0.34]{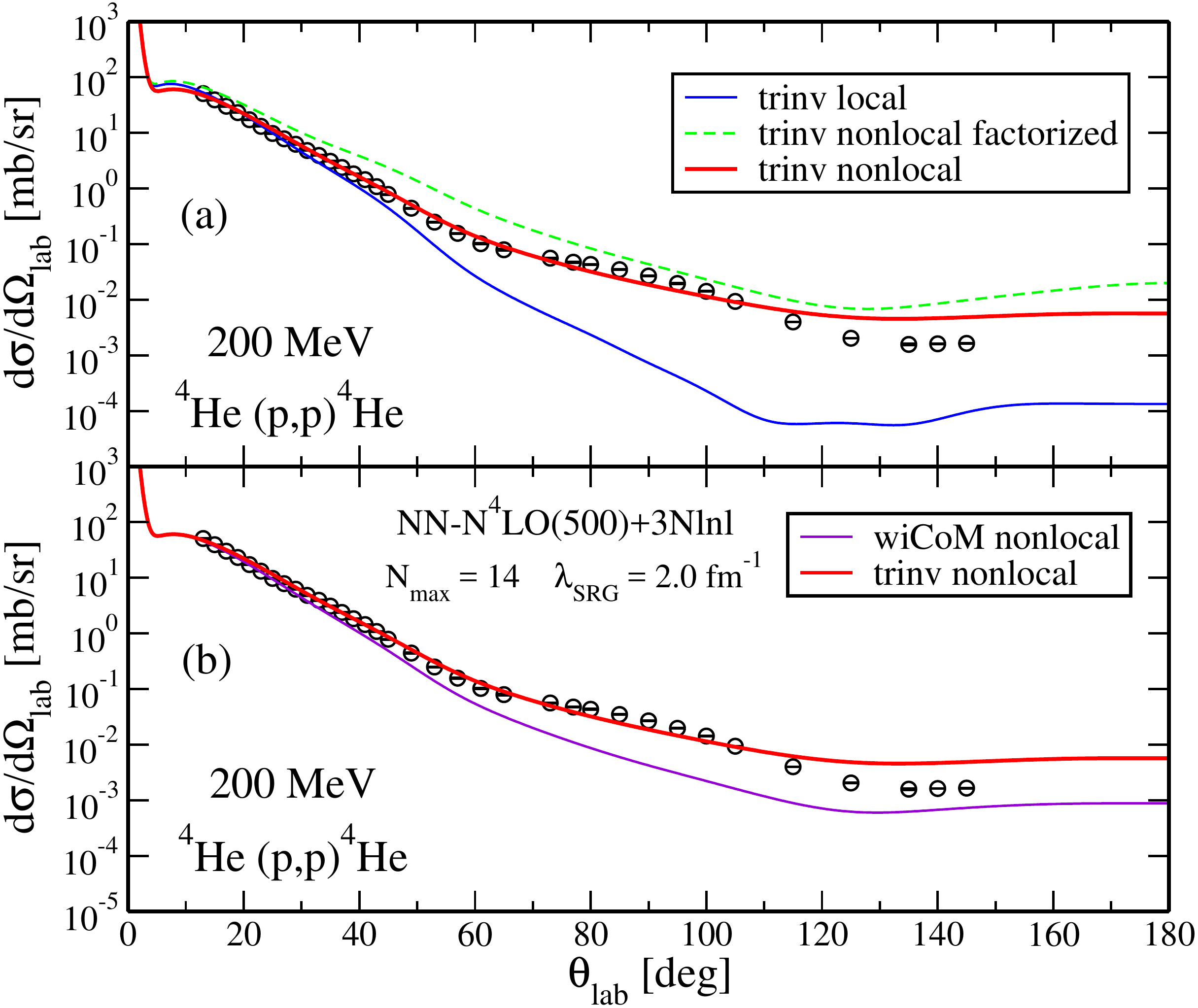}
\caption{\label{4He_local_nonlocal} (Color online) {\bf Panel (a):} differential cross sections as functions of the laboratory scattering angle computed from {\it trinv} local and
nonlocal densities. The dashed line is obtained from Eq.~(\ref{factorizedop}) with the density profile computed using Eq.~(\ref{density_profile}).
{\bf Panel (b):} differential cross sections as functions of the laboratory scattering angle computed from {\it wiCOM} and {\it trinv} nonlocal densities.
All cross sections were computed for the \textsuperscript{4}He(p,p)\textsuperscript{4}He reaction at $200$ MeV (laboratory energy) and using
the bare NN-N\textsuperscript{4}LO(500) interaction for the calculation of the free $NN$ $t$ matrix. All densities were computed with the
NN-N\textsuperscript{4}LO(500)+3Nlnl interaction with $\nmax = 14$, $\hb = 20$ MeV, and $\lambda_{\mathrm{SRG}} = 2.0$ fm\textsuperscript{-1}.
Experimental data are taken from Ref.~\cite{PhysRevC.21.1932}.}
\end{center}
\end{figure}

\begin{figure}[t]
\begin{center}
\includegraphics[scale=0.34]{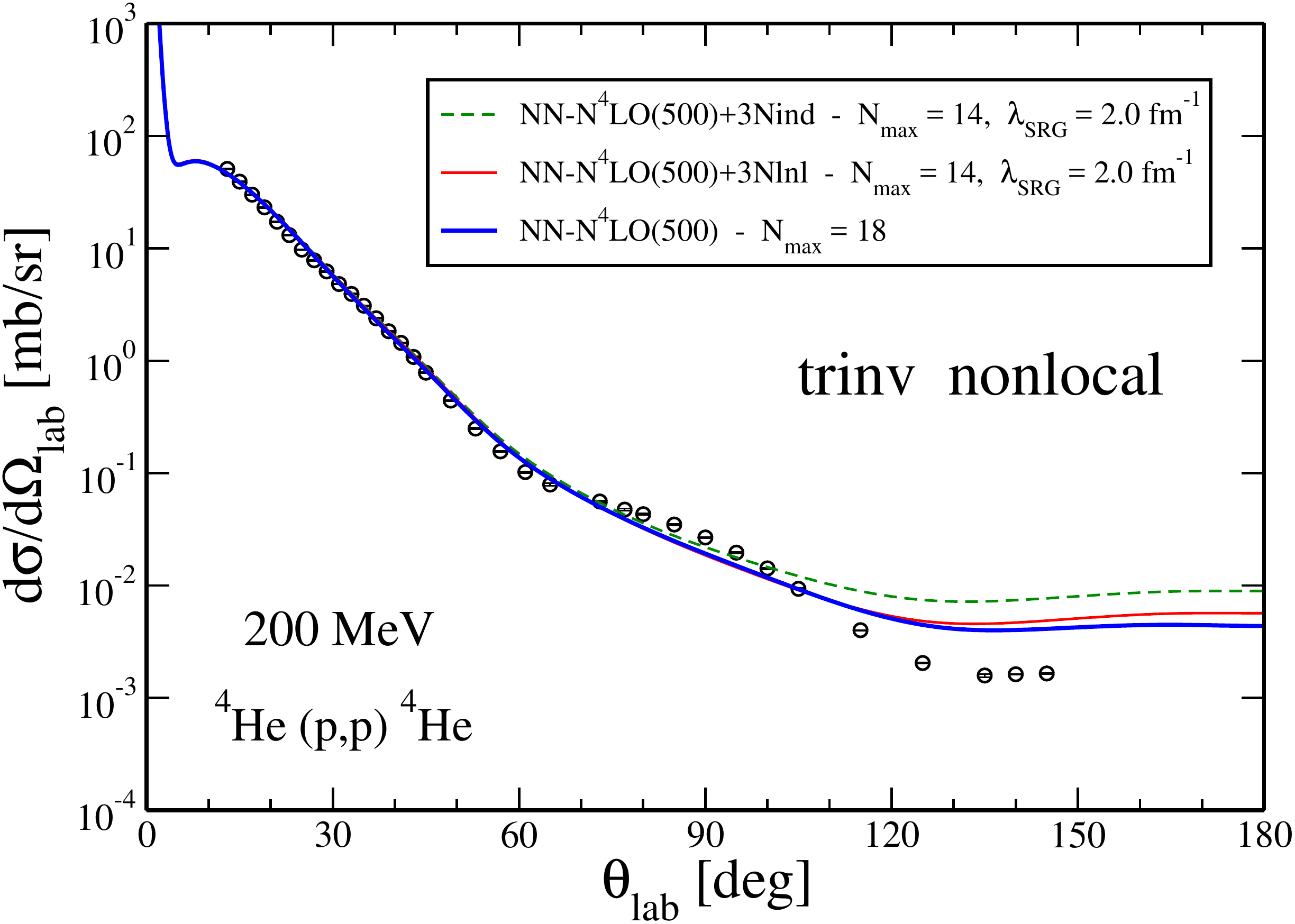}
\caption{\label{4He_sigma_hw} (Color online) Differential cross sections as functions of the laboratory scattering angle computed from {\it trinv} nonlocal densities
for the \textsuperscript{4}He(p,p)\textsuperscript{4}He reaction at $200$ MeV (laboratory energy) and using the bare NN-N\textsuperscript{4}LO(500) interaction for the calculation of
the free $NN$ $t$ matrix. The densities were computed with the SRG-evolved NN-N\textsuperscript{4}LO(500) interaction plus the three-body SRG-induced (3Nind) or the full
three-body (3Nlnl) interaction in the first two cases, while we used the only bare NN-N\textsuperscript{4}LO(500) interaction in the third case. In the first two cases the results
were obtained with $\nmax = 14$, $\hb = 20$ MeV, and $\lambda_{\mathrm{SRG}} = 2.0$ fm\textsuperscript{-1}, while in the third case we used $\nmax = 18$ and $\hb = 20$ MeV.
Experimental data are taken from Ref.~\cite{PhysRevC.21.1932}.}
\end{center}
\end{figure}

\begin{figure}[t]
\begin{center}
\includegraphics[scale=0.34]{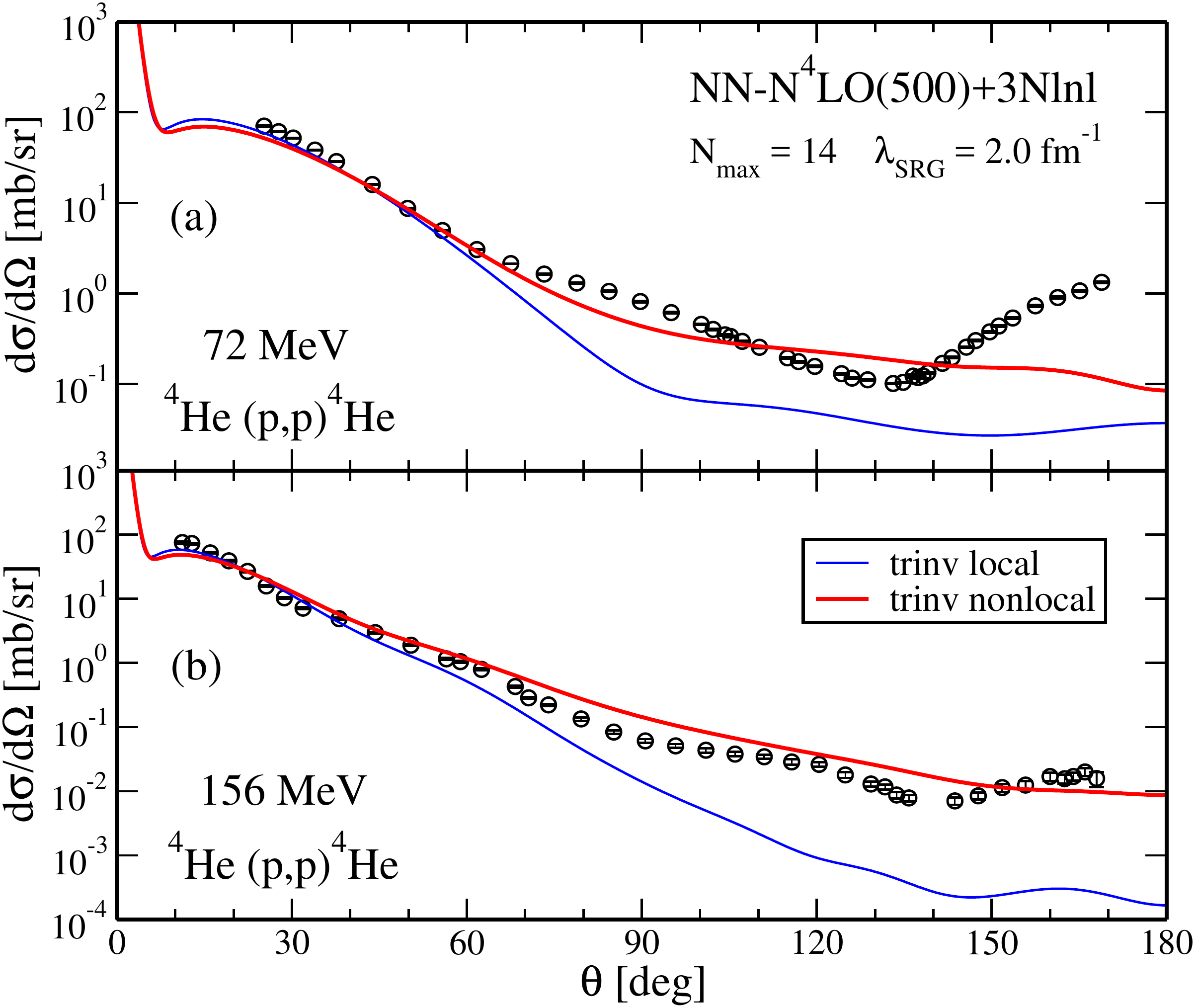}
\caption{\label{4He_72_156MeV_sigma} (Color online) Differential cross sections in the $p A$ center-of-mass frame computed from {\it trinv} local and nonlocal densities
for the \textsuperscript{4}He(p,p)\textsuperscript{4}He reaction at the incident proton energy in the laboratory frame of $72$ MeV (a) and $156$ MeV (b), respectively.
For all calculations the density was computed with the SRG-evolved NN-N\textsuperscript{4}LO(500)+3Nlnl interaction at $\nmax = 14$ and with $\hb = 20$ MeV
and $\lambda_{\mathrm{SRG}} = 2.0$ fm\textsuperscript{-1}, while the free $NN$ $t$ matrix was computed with the bare NN-N\textsuperscript{4}LO(500) interacton.
Experimental data are taken from Refs.~\cite{PhysRevC.39.56,PhysRevC.12.251}.}
\end{center}
\end{figure}

\begin{figure}[t]
\begin{center}
\includegraphics[scale=0.34]{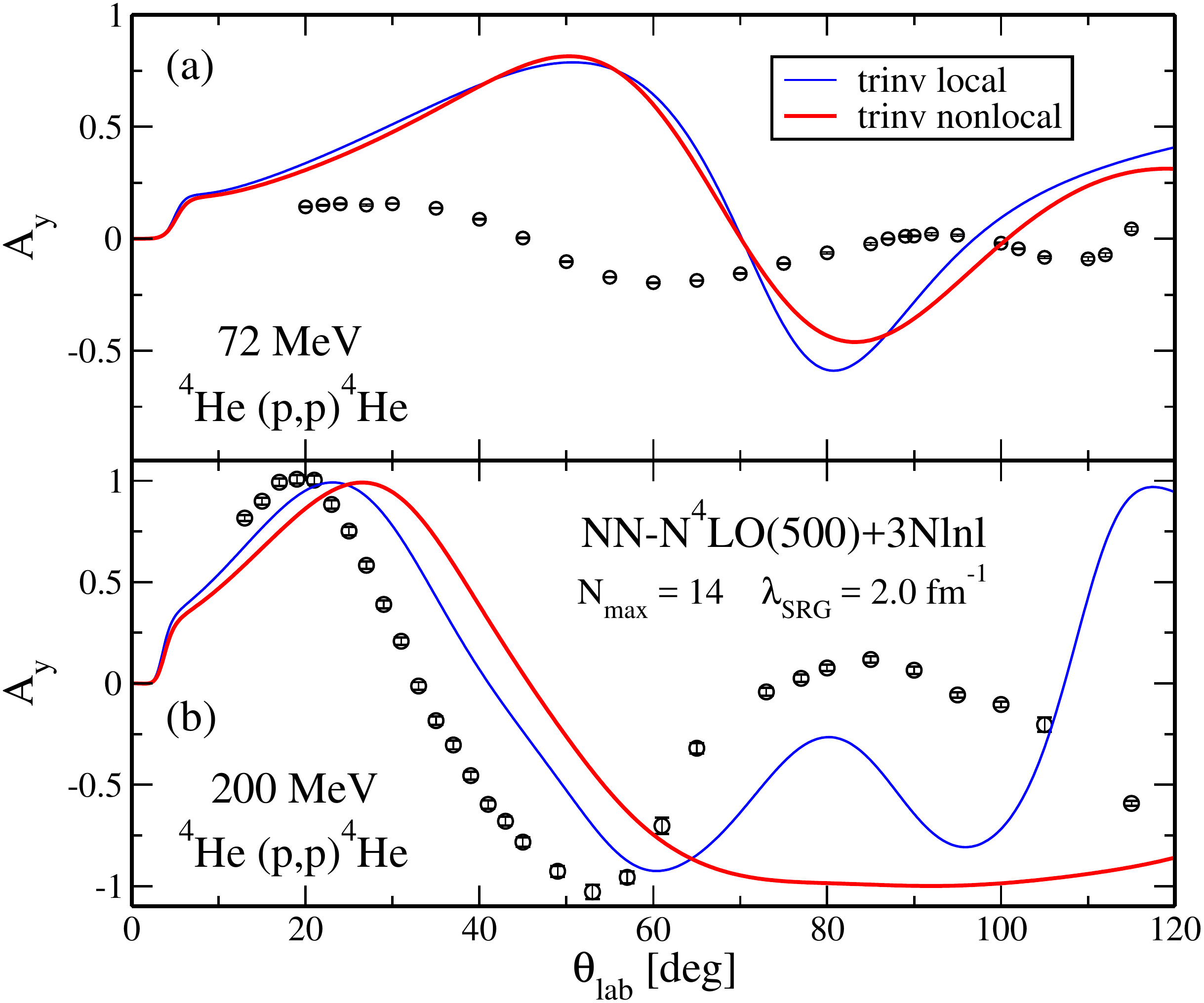}
\caption{\label{4He_72_156MeV_ay} (Color online) The same as in Fig.~\ref{4He_72_156MeV_sigma} but for the analyzing power at the incident proton energy in the
laboratory frame of $72$ MeV (a) and $200$ MeV (b), respectively.
Experimental data are taken from Refs.~\cite{PhysRevC.39.56,PhysRevC.21.1932}.}
\end{center}
\end{figure}

\begin{figure}[t]
\begin{center}
\includegraphics[scale=0.34]{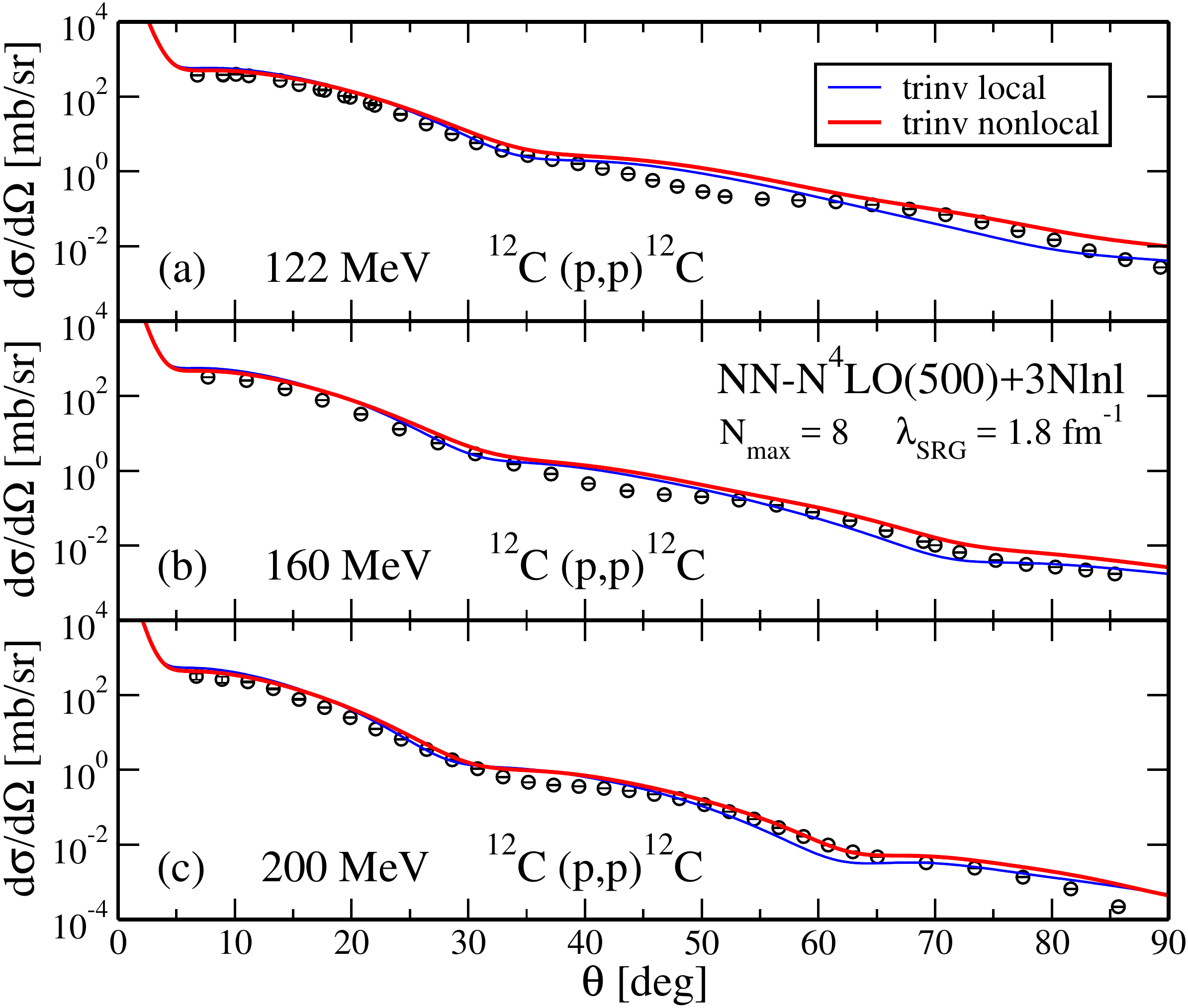}
\caption{\label{12C_sigma} (Color online) Differential cross sections in the $p A$ center-of-mass frame computed from {\it trinv} local and nonlocal densities
for the \textsuperscript{12}C(p,p)\textsuperscript{12}C reaction at the incident proton energy in the laboratory frame of $122$ (a), $160$ (b), and $200$ MeV (c), respectively.
For all calculations the densities were computed with the SRG-evolved NN-N\textsuperscript{4}LO(500)+3Nlnl interaction at $\nmax = 8$ and with $\hb = 20$ MeV
and $\lambda_{\mathrm{SRG}} = 1.8$ fm\textsuperscript{-1}, while the free $NN$ $t$ matrix was computed with the bare NN-N\textsuperscript{4}LO(500) interaction.
Experimental data are taken from Ref.~\cite{exfor}.}
\end{center}
\end{figure}

\begin{figure}[t]
\begin{center}
\includegraphics[scale=0.34]{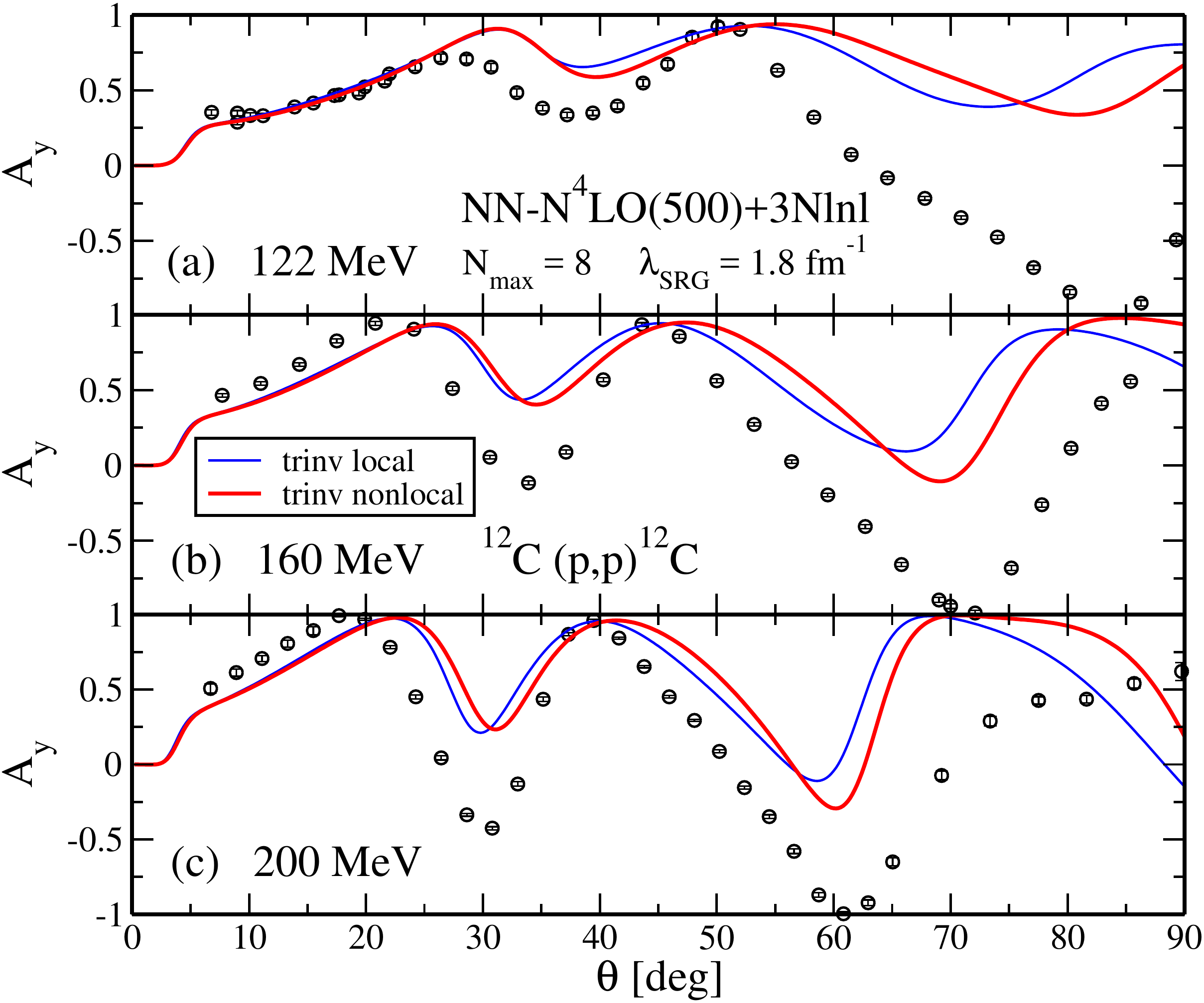}
\caption{\label{12C_Ay} (Color online) The same as in Fig.~\ref{12C_sigma} but for the analyzing power.}
\end{center}
\end{figure}

\begin{figure}[t]
\begin{center}
\includegraphics[scale=0.34]{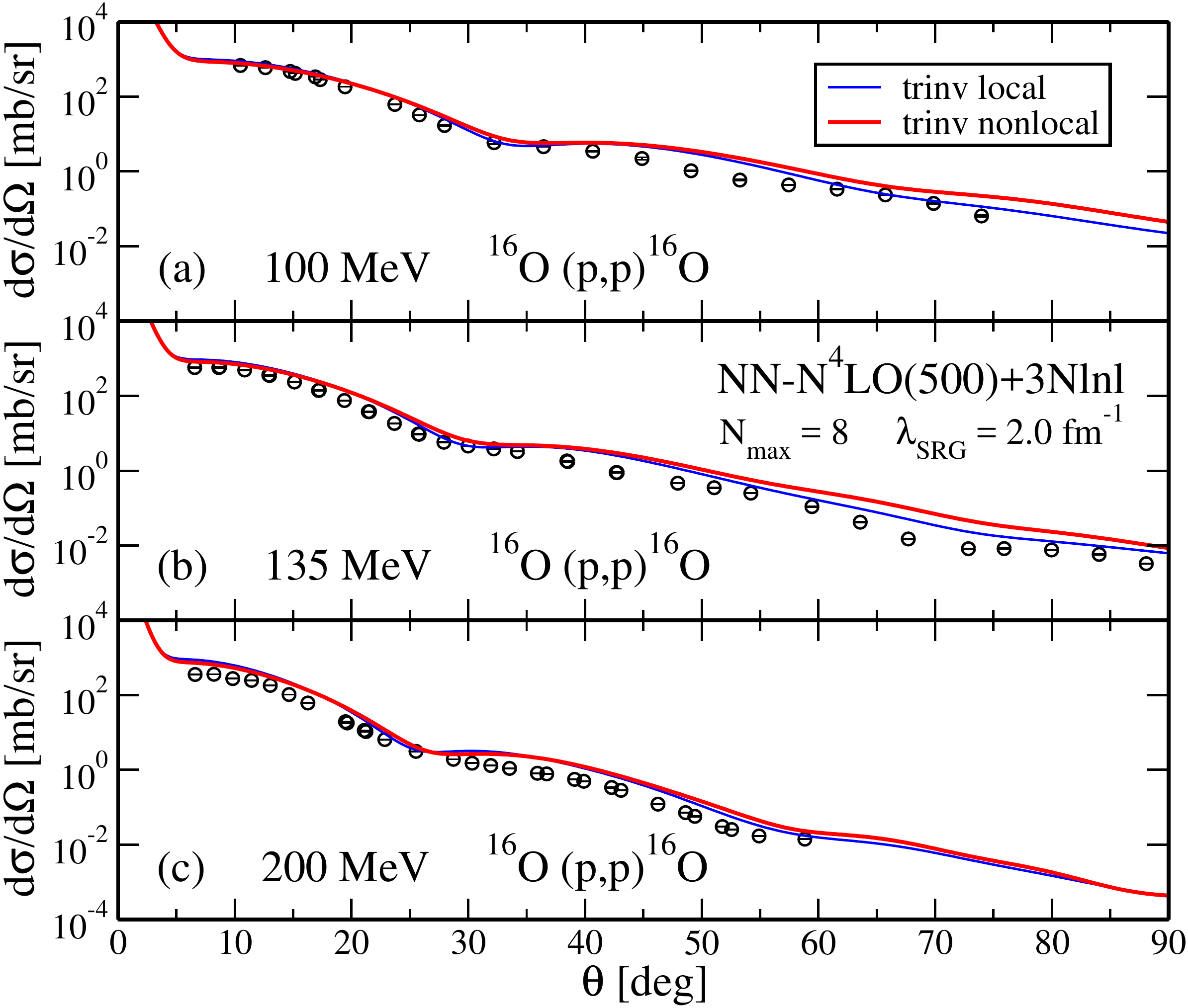}
\caption{\label{16O_sigma} (Color online) The same as in Fig.~\ref{12C_sigma} but for the \textsuperscript{16}O(p,p)\textsuperscript{16}O reaction at the laboratory
energy of $100$ (a), $135$ (b), and $200$ MeV (c), respectively.
Experimental data are taken from Ref.~\cite{kelly}.}
\end{center}
\end{figure}

\begin{figure}[t]
\begin{center}
\includegraphics[scale=0.34]{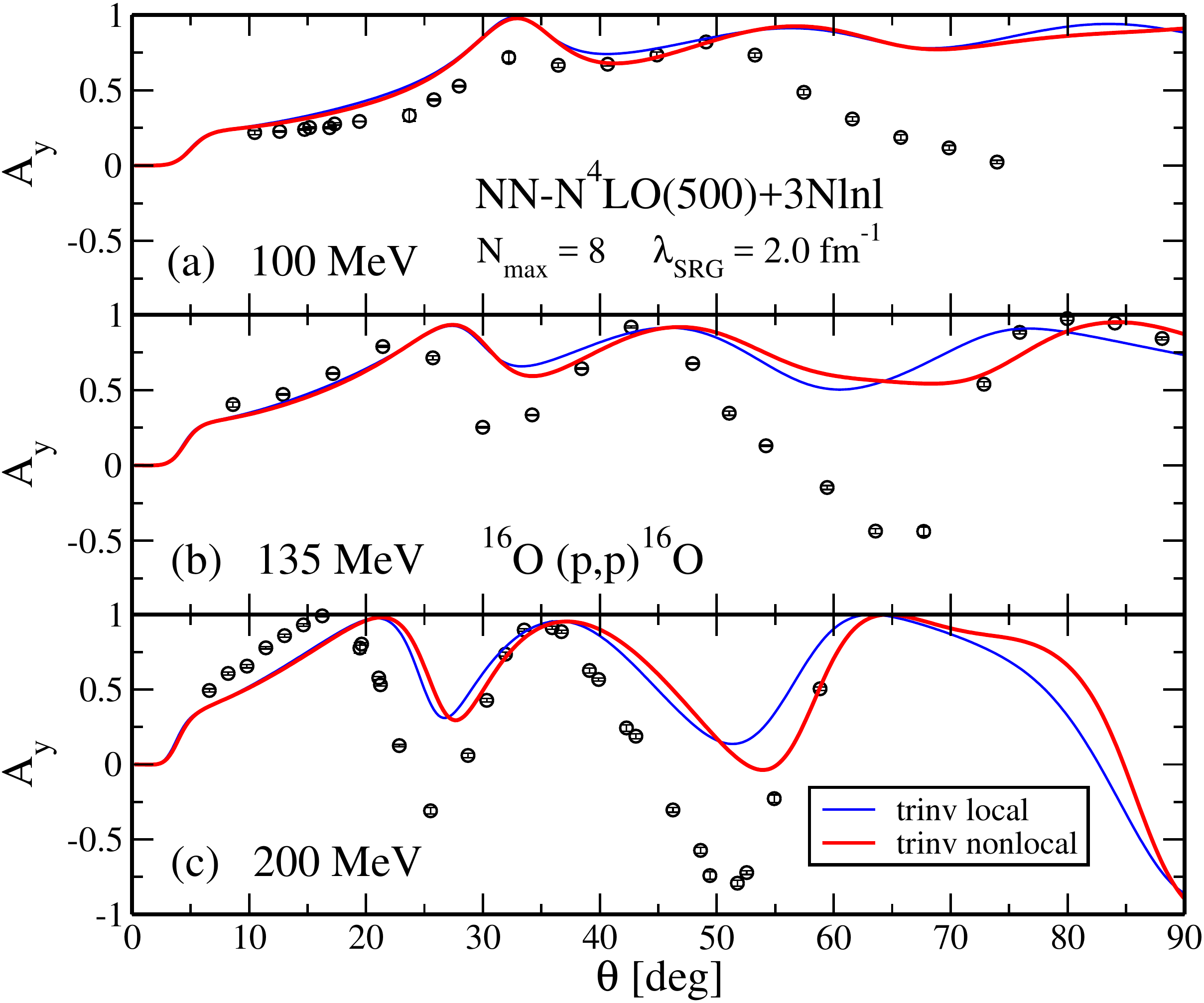}
\caption{\label{16O_Ay} (Color online) The same as in Fig.~\ref{16O_sigma} but for the analyzing power.}
\end{center}
\end{figure}

\begin{figure}[t]
\begin{center}
\includegraphics[scale=0.34]{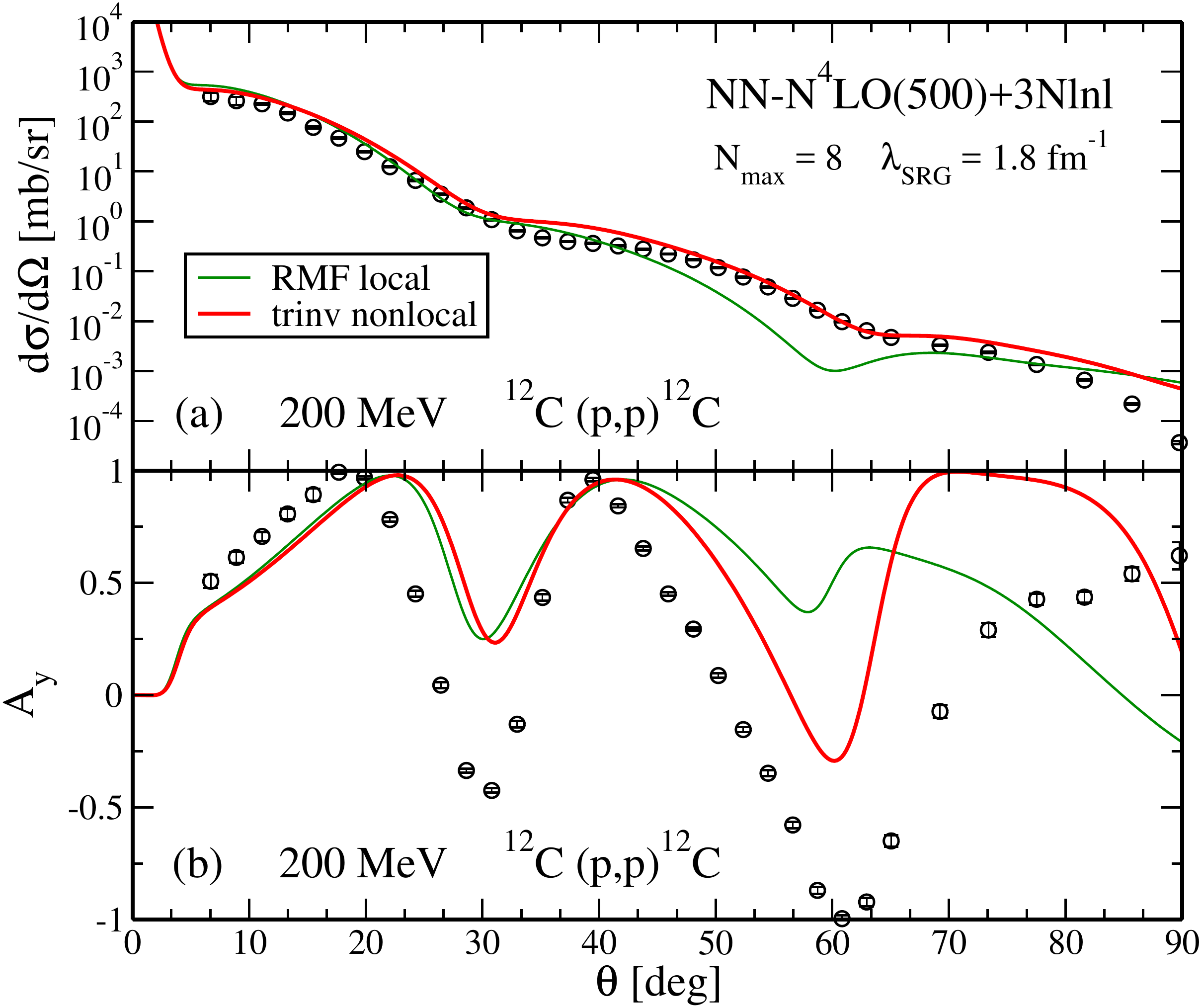}
\caption{\label{12C_comp} (Color online) Differential cross section (a) and analyzing power (b) in the $p A$ center-of-mass frame computed from local RMF
and {\it trinv} nonlocal densities for the \textsuperscript{12}C(p,p)\textsuperscript{12}C reaction at the incident proton energy in the laboratory frame of $200$ MeV.
In all calculations the free $NN$ $t$ matrix was computed with the bare NN-N\textsuperscript{4}LO(500) interaction, while the {\it trinv} nonlocal density was computed with the
NN-N\textsuperscript{4}LO(500)+3Nlnl interaction. Experimental data are taken from Ref.~\cite{exfor}.}
\end{center}
\end{figure}

\begin{figure}[t]
\begin{center}
\includegraphics[scale=0.34]{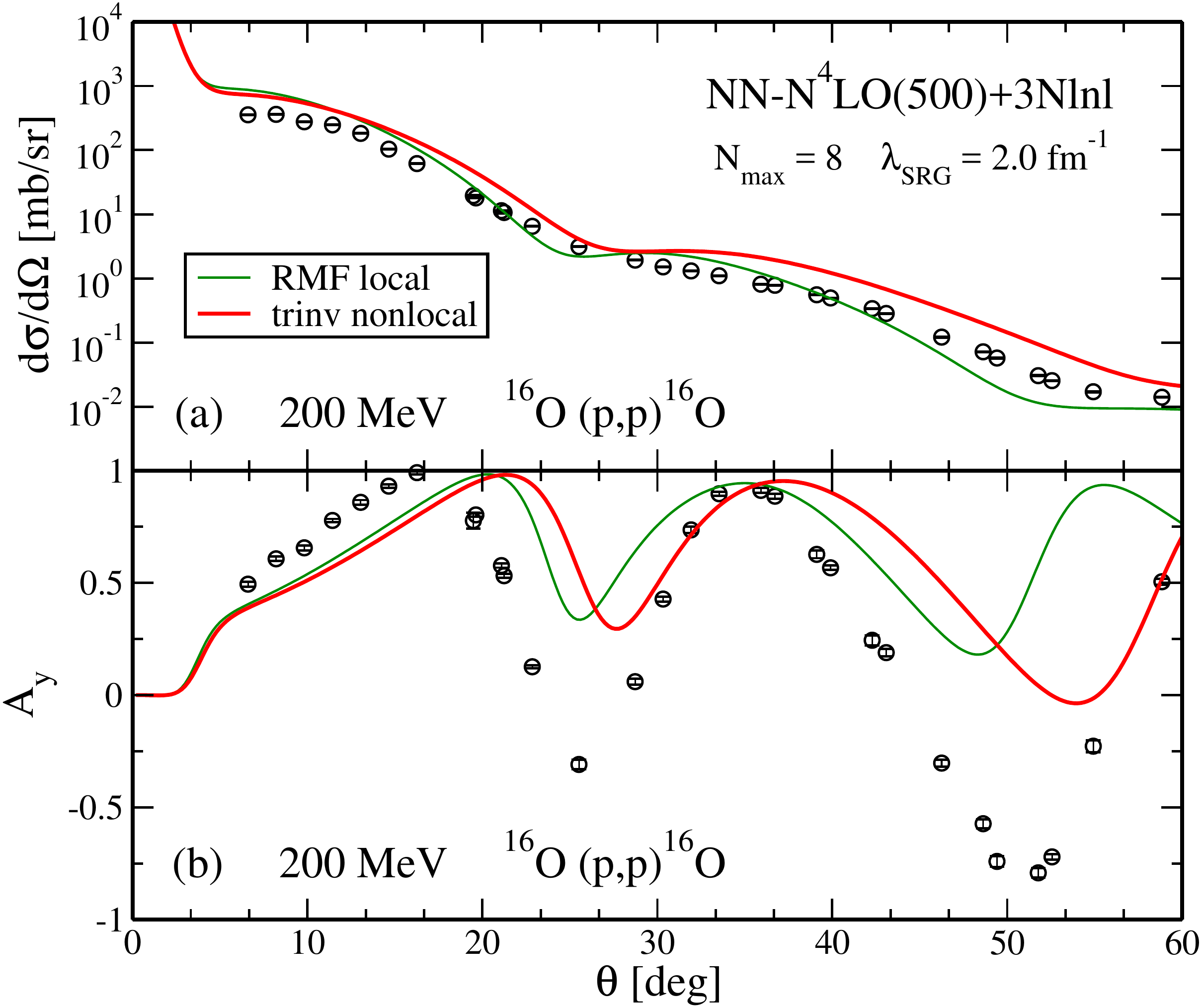}
\caption{\label{16O_comp} (Color online) The same as in Fig.~\ref{12C_comp} but for the \textsuperscript{16}O(p,p)\textsuperscript{16}O reaction.
Experimental data are taken from Ref.~\cite{kelly}.}
\end{center}
\end{figure}

Since the densities are one of the two basic ingredients for the calculation of the optical potential, we start our discussion from the neutron and proton nuclear density profiles.
In Fig.~\ref{density_profiles} we show the results for $\rho_{\alpha} (q)$, calculated for \textsuperscript{4}He, \textsuperscript{12}C, and \textsuperscript{16}O, using {\it trinv} local
and nonlocal densities. The results obtained from nonlocal densities were computed using Eq.~(\ref{density_profile}), while the results for the local ones were
computed using Eq.~(\ref{ftlocaldens}). Moreover, for \textsuperscript{12}C and \textsuperscript{16}O we also show the nuclear density
profiles computed within the Relativistic Mean-Field (RMF) description~\cite{Niksic20141808,PhysRevC.66.024306} of spherical nuclei. As can be seen, the profiles obtained from
nonlocal densities exhibit a larger extent toward higher values of momentum transfer and the largest difference is obtained for \textsuperscript{4}He, that is thus chosen to analyze
the differences between local and nonlocal, {\it wiCOM} and {\it trinv} calculations.

In panel (a) of Fig.~\ref{4He_local_nonlocal} we present the results for the differential cross section of proton elastic scattering off \textsuperscript{4}He obtained using {\it trinv}
local and nonlocal densities, computed with the NN-N\textsuperscript{4}LO(500)+3Nlnl interaction. The cross section computed with the {\it trinv} nonlocal density (thick
solid line) was obtained using Eq.~(\ref{fullfoldingop}) while the cross section computed with the {\it trinv} local density (thin solid line) was calculated using
Eq.~(\ref{factorizedop}). As can be seen, the advantage of using these new nonlocal densities is evident. Although the two calculations give basically the same results for angles
smaller than $40$ degrees, after this value the difference between the two curves starts to increase and around $100$ degrees the discrepancy is about two orders of magnitude.
The result computed with the nonlocal density is able to describe the experimental data, showing a good agreement up to $100$ degrees and then slightly overestimating them.
In order to assess the benefit produced by using Eq.~(\ref{fullfoldingop}), we also show the result (dashed line) obtained from Eq.~(\ref{factorizedop}) but with the density
profile computed using Eq.~(\ref{density_profile}) with the {\it trinv} nonlocal density. Here the only difference between this result and that one displayed with the thick solid
line is that, in the second case, the folding integral has been performed. As can been seen the two curves display the same trend and are close to each other, although the dashed
line always overestimates the experimental data. This result is not surprising and it is a consequence of the different behaviour of the density profiles displayed in
Fig.~\ref{density_profiles}. Instead, in panel (b) of Fig.~\ref{4He_local_nonlocal} we compare the results for the differential cross section obtained with {\it trinv} (thick solid line)
and {\it wiCOM} (thin solid line) nonlocal densities. Since the effect of the COM contamination diminishes with $A$, \textsuperscript{4}He is still the best candidate to analyze
the effect produced on the scattering observables, as it will be the largest. In fact, we can see that in this case the discrepancy between the two results starts to
increase beyond $40$ degrees and the COM contamination in the density produces a cross section that differs from that one obtained with the {\it trinv} density by one order of
magnitude. We also computed the scattering observables with {\it wiCOM} densities for \textsuperscript{12}C and \textsuperscript{16}O and, as expected, we found a smaller but
non-negligible effect.

Before we show the scattering observables for other energies and for other nuclei we want to discuss the consistency of our calculations. As outlined in
Sec.~\ref{sec_optical_potential}, the model for the optical potential only includes the two-body $NN$ interaction and thus is not completely consistent with the densities, that were
obtained with two- and three-body interactions. Moreover, the calculation of the densities was performed with the SRG-evolved $NN$ potential, while for the calculation of the
$t$ matrix we used the bare $NN$ interaction. Unfortunately, the inclusion of three-nucleon forces in the optical potential is a difficult task and it cannot be performed
at the moment. In principle, fully consistent results can be produced if only the $NN$ bare interaction is used to compute the density and the $t$ matrix, but, in practice, this procedure
is feasible only for very light nuclei, like \textsuperscript{4}He. In fact, the three-body interaction is important to produce realistic binding energies and radii, and the SRG
evolution of the $NN$ potential is necessary to obtain converged results at smaller model space sizes. This is particularly important for nuclei like \textsuperscript{12}C and
\textsuperscript{16}O as they require large model spaces, which prohibits the calculation of the nuclear wave functions with only the $NN$ bare interaction.
In Fig.~\ref{4He_sigma_hw} we consider the \textsuperscript{4}He(p,p)\textsuperscript{4}He reaction at $200$ MeV in the laboratory frame and we display
the differential cross section computed with three different {\it trinv} nonlocal densities. In the first case (dashed line) the density was obtained with the SRG-evolved
NN-N\textsuperscript{4}LO(500) potential plus the SRG-induced three-nucleon forces (3Nind), in the second case (thin solid line) we used the SRG-evolved
NN-N\textsuperscript{4}LO(500) potential plus the full three-nucleon forces (3Nlnl), as shown in Fig.~\ref{4He_local_nonlocal}, and in the third case (thick solid line) the
density was computed with only the bare NN-N\textsuperscript{4}LO(500) potential and it thus represents the fully consistent calculation. The result obtained with only the $NN$
potential reproduces the experimental data very well up to $100$ degrees and it is in good agreement with the other curves, especially with the one obtained with the full three-nucleon
interaction. Here, it is important to notice that to obtain a converged density with the only NN-N\textsuperscript{4}LO(500) bare potential we performed our calculations up
to $\nmax = 18$. That specifies a much larger model space than one used for the other densities. We also investigated the $\hb$ dependence of the results obtained with
the bare $NN$ potential, performing our calculations for $\hb = 20$, $24$, and $28$. In all these cases our findings were very close to each other with a slightly appreciable
difference only at very large scattering angles.

In Fig.~\ref{4He_72_156MeV_sigma} we show the differential cross section for the same reaction and the same interaction considered above, but computed at $72$
and $156$ MeV, respectively. Similarly, for these lower energies the calculations performed with {\it trinv} nonlocal densities provide a better description of the experimental data.
At $156$ MeV the overall shape of the cross section is reproduced and there is a good agreement with the data for angles up to $70$ degrees, while for larger
angles they are somewhat overestimated. We also see that for large angles the difference between the two theoretical results is about two orders of magnitude.
At $72$ MeV the situation is a bit different. In fact, we remind the reader that the expression for the optical potential was obtained in the impulse approximation, which consists of
neglecting the coupling between the struck target nucleon and the residual nucleus. This is a good approximation for energies around $200$ MeV and beyond, but
it cannot provide a good description of the experimental data at lower energies, where the medium effects are more important. We also see that in this case
the difference between the two theoretical curves persists and the nonlocal density provides an overall reasonable description of the data with a good
agreement for angles up to $60$ degrees.
 
For the sake of completeness, in Fig.~\ref{4He_72_156MeV_ay} we show the results for the analyzing power ($A_y$) for energies where experimental data exists.
Reproducing $A_y$ is always difficult since this observable is more sensitive than the differential cross section and a good description of the data can only be achieved
with the proper inclusion of effects that are missing in the current calculations. Of course, it is not surprising that the data at $72$ MeV are not reproduced well, and this
can be easily explained by the lack of medium effects due to the impulse approximation. At $200$ MeV the theoretical results match data for small angles
but they seem to be shifted toward larger angles. Beyond $60$ degrees, $A_y$ obtained with the local density exhibits a local maximum and then
it displays an unphysical upward trend, while the result obtained with the nonlocal density reaches an almost constant value.

In Fig.~\ref{12C_sigma} we show the results for the differential cross section of proton elastic scattering off \textsuperscript{12}C computed with the {\it trinv} local and nonlocal
densities at the three laboratory energies of $122$, $160$, and $200$ MeV, respectively. Also in this case the results were obtained using the
NN-N\textsuperscript{4}LO(500)+3Nlnl interaction for the calculation of the densities and with the NN-N\textsuperscript{4}LO(500) interaction for the $NN$ $t$ matrix.
For this target nucleus we see that the difference between the results obtained with local and nonlocal densities is much smaller than the one obtained for
\textsuperscript{4}He, and both curves are in very good agreement with the experimental data, especially at $200$ MeV, which is a proper energy for the impulse
approximation. In particular, for this energy the curve computed with the nonlocal density gives a good description of the experimental data in the region of the
second minimum, while for all three energies the agreement between theoretical results and data is a bit less accurate in a region of about $10$ degrees of width after the first minimum.

Additionally, for \textsuperscript{12}C we show the calculations for the analyzing power in Fig.~\ref{12C_Ay}. For all energies the theoretical results give a reasonable description
of the experimental data, reproducing their general shape, but the agreement with them is poor, especially at larger angles and lower energies. The calculations performed with
the local density seem to be shifted toward smaller angles with respect the values of the two minima of the data, that are instead better reproduced by the nonlocal densities.

In Fig.~\ref{16O_sigma} and Fig.~\ref{16O_Ay} we show the differential cross section and the analyzing power for the \textsuperscript{16}O(p,p)\textsuperscript{16}O reaction,
computed at the laboratory energies of $100$, $135$, and $200$ MeV, respectively. Of course, the calculations have been performed with the same interaction used
for \textsuperscript{12}C. In this case the results with local and nonlocal densities are close to each other and the experimental data of the differential cross section
are globally well reproduced by our calculations, although after the first minimum they are slightly overestimated. Concerning the $A_y$, the theoretical curves display
a reasonable shape, but, again, the agreement with the data is poor near minima.

Finally, in order to make a comparison with other methods, we generated local densities for \textsuperscript{12}C and \textsuperscript{16}O using a RMF
approach~\cite{Niksic20141808,PhysRevC.66.024306}, the same employed in our previous works~\cite{PhysRevC.93.034619,PhysRevC.96.044001}.
In Fig.~\ref{12C_comp} we show the results for the differential cross section and the analyzing power for the \textsuperscript{12}C(p,p)\textsuperscript{12}C reaction at
$200$ MeV, obtained with our new {\it trinv} nonlocal density, and we compare them with those obtained using Eq.~(\ref{factorizedop}) with a RMF local density.
For the cross section, the two curves give very similar results and are in agreement with the experimental data, but in the region of the second minimum the result
obtained with the RMF density underestimates the data. The same behaviour is reflected in the $A_y$, where the general shape of the second minimum is not reproduced.
In Fig.~\ref{16O_comp} we show the same calculations for the \textsuperscript{16}O(p,p)\textsuperscript{16}O reaction. In this case the situation is a bit different.
The general description of the data is good; at large angles the differential cross section is underestimated by the curve obtained with the RMF density, while
it is overestimated by that one obtained with the {\it trinv} nonlocal density. The main difference with respect to the results of Fig.~\ref{12C_comp} is that the curve
corresponding to the {\it trinv} density seems to be shifted toward larger angles; this is particularly evident for the analyzing power. This behaviour could be partially explained
looking at Fig.~\ref{density_profiles}, which shows that for \textsuperscript{12}C the RMF profile is close to the {\it trinv} nonlocal one and equal to the profile obtained with the {\it trinv}
local density. This is not true for \textsuperscript{16}O where the difference is larger.

\subsection{Results for \textsuperscript{6,8}He}
\label{sec_halo_nuclei}

\begin{figure}[t]
\begin{center}
\includegraphics[scale=0.32]{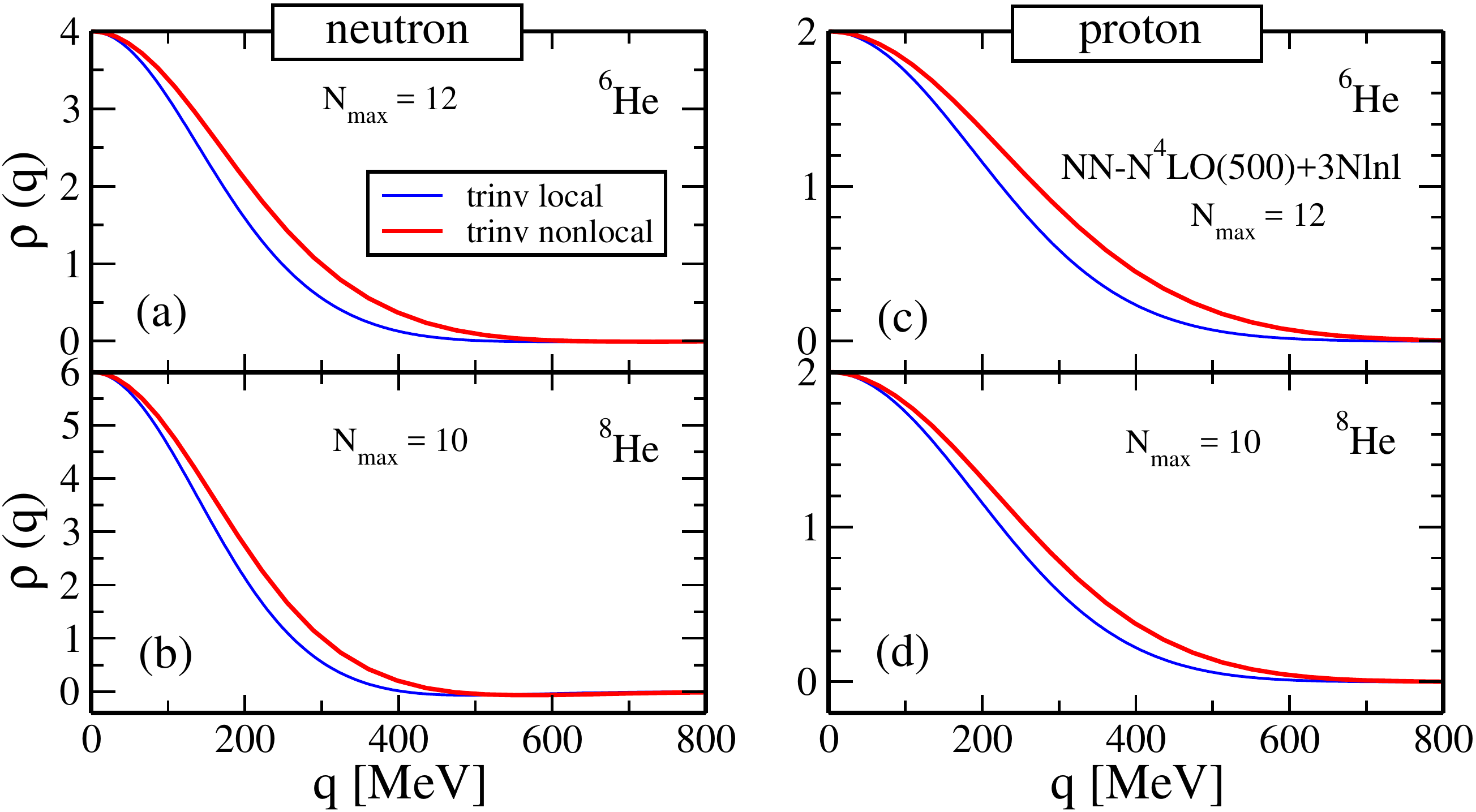}
\caption{\label{density_profiles_halo} (Color online) The same as in Fig.~\ref{density_profiles} but for \textsuperscript{6,8}He.}
\end{center}
\end{figure}

\begin{figure}[t]
\begin{center}
\includegraphics[scale=0.32]{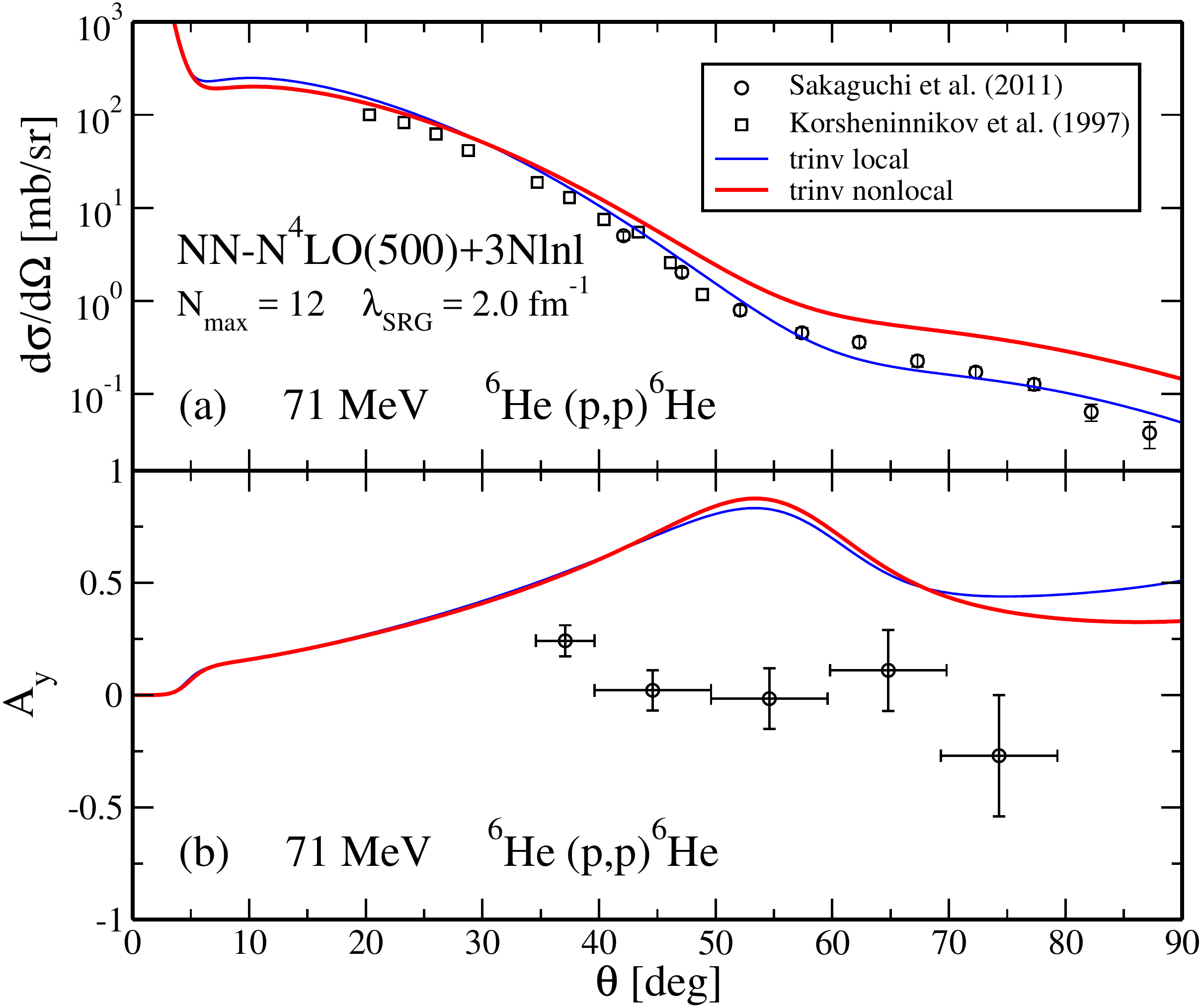}
\caption{\label{6He_71MeV} (Color online) Differential cross section (a) and analyzing power (b) in the $p A$ center-of-mass frame computed from {\it trinv} local and nonlocal
densities for elastic scattering of \textsuperscript{6}He at the projectile energy of $71$ MeV/nucleon. For all calculations the densities were computed with the SRG-evolved
NN-N\textsuperscript{4}LO(500)+3Nlnl interaction at $\nmax = 12$ and with $\hb = 20$ MeV and $\lambda_{\mathrm{SRG}} = 2.0$ fm\textsuperscript{-1}, while the free
$NN$ $t$ matrix was computed with the bare NN-N\textsuperscript{4}LO(500) interaction.
Experimental data are taken from Ref.~\cite{PhysRevC.84.024604,KORSHENINNIKOV199745}.}
\end{center}
\end{figure}

\begin{figure}[t]
\begin{center}
\includegraphics[scale=0.32]{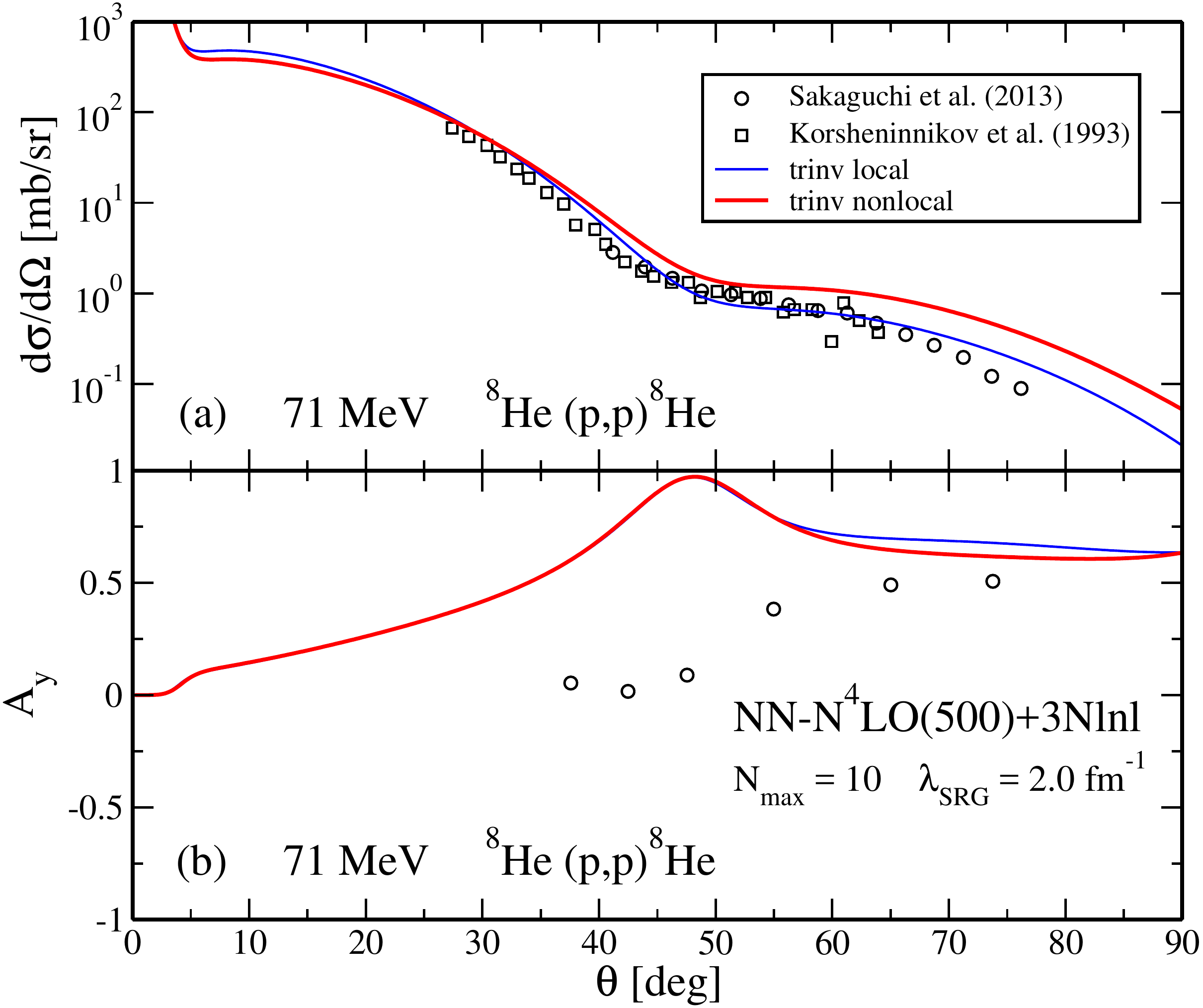}
\caption{\label{8He_71MeV} (Color online) The same as in Fig.~\ref{6He_71MeV} but for \textsuperscript{8}He and $\nmax = 10$.
Experimental data are taken from Ref.~\cite{PhysRevC.87.021601,KORSHENINNIKOV199338}.}
\end{center}
\end{figure}

\begin{figure}[t]
\begin{center}
\includegraphics[scale=0.32]{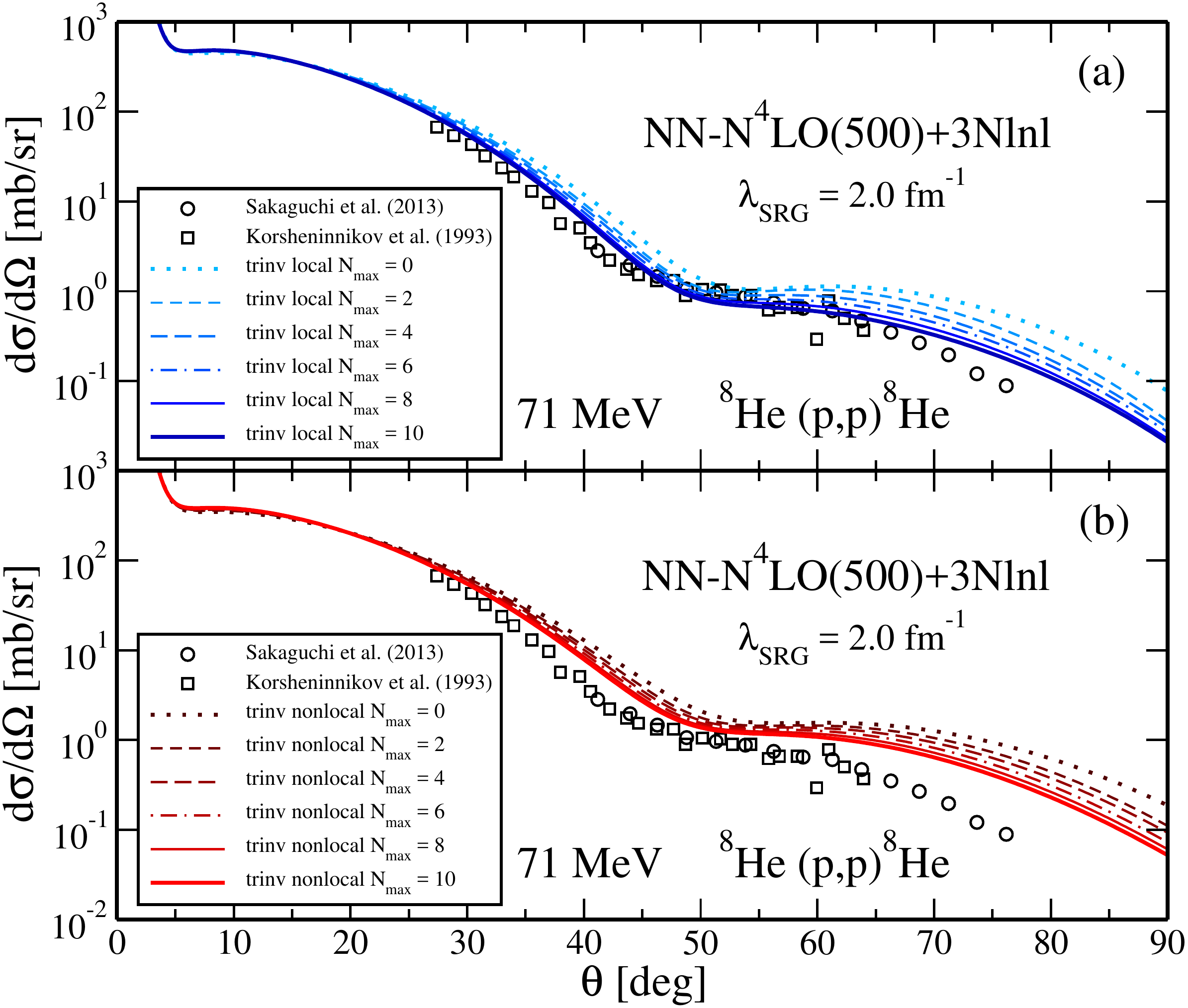}
\caption{\label{8He_conv} (Color online) Convergence pattern for the differential cross section in the $p A$ center-of-mass frame computed from {\it trinv} local (a) and
nonlocal (b) densities for elastic scattering of \textsuperscript{8}He at the projectile energy of $71$ MeV/nucleon. For all calculations the densities were computed with the
SRG-evolved NN-N\textsuperscript{4}LO(500)+3Nlnl interaction at $\nmax = 10$ and with $\hb = 20$ MeV and $\lambda_{\mathrm{SRG}} = 2.0$ fm\textsuperscript{-1}, while the
free $NN$ $t$ matrix was computed with the bare NN-N\textsuperscript{4}LO(500) interaction.
Experimental data are taken from Ref.~\cite{PhysRevC.87.021601,KORSHENINNIKOV199338}.}
\end{center}
\end{figure}

\begin{figure}[t]
\begin{center}
\includegraphics[scale=0.32]{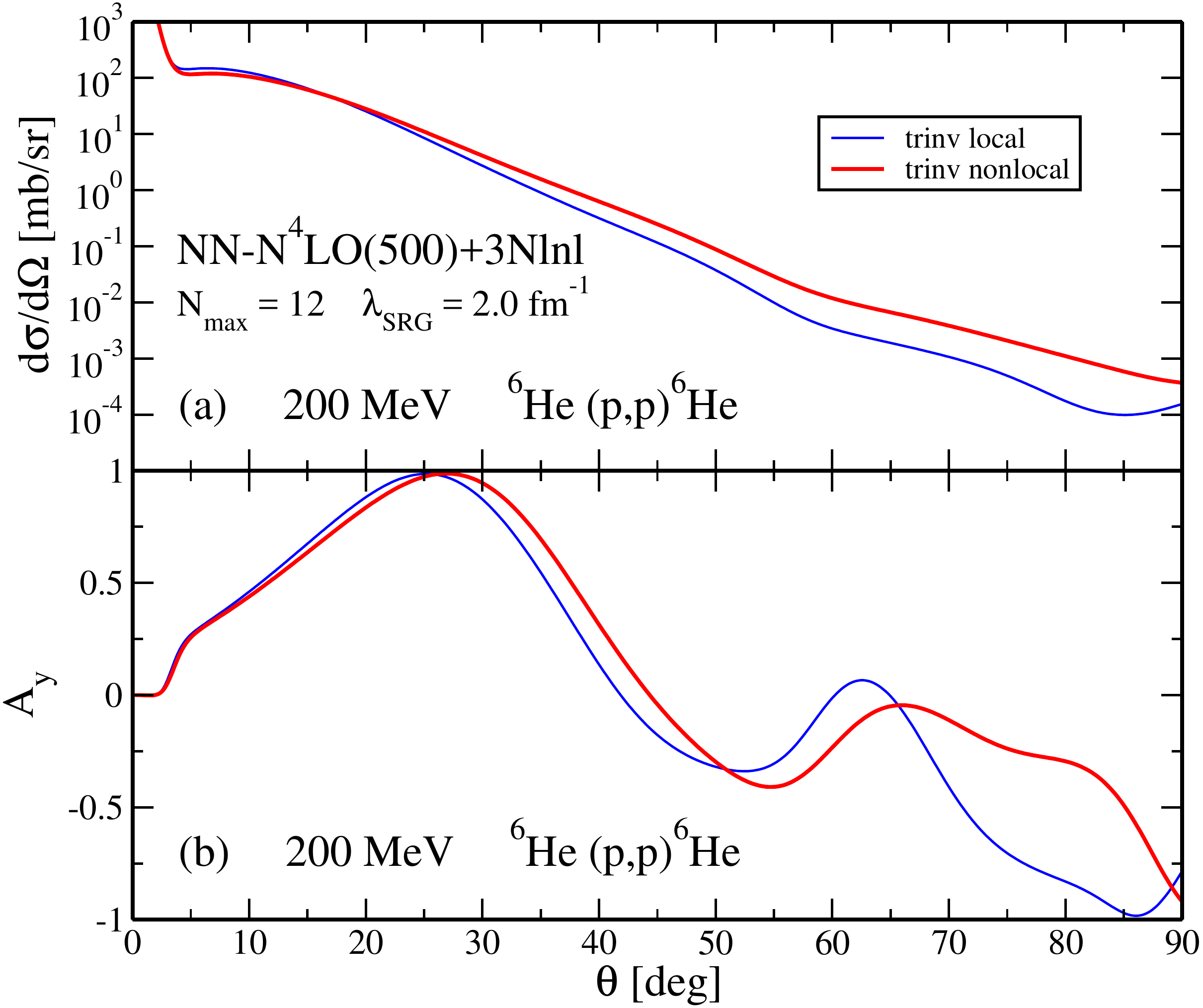}
\caption{\label{6He_200MeV} (Color online) The same as in Fig.~\ref{6He_71MeV} but at the projectile energy of $200$ MeV/nucleon.}
\end{center}
\end{figure}

\begin{figure}[t]
\begin{center}
\includegraphics[scale=0.32]{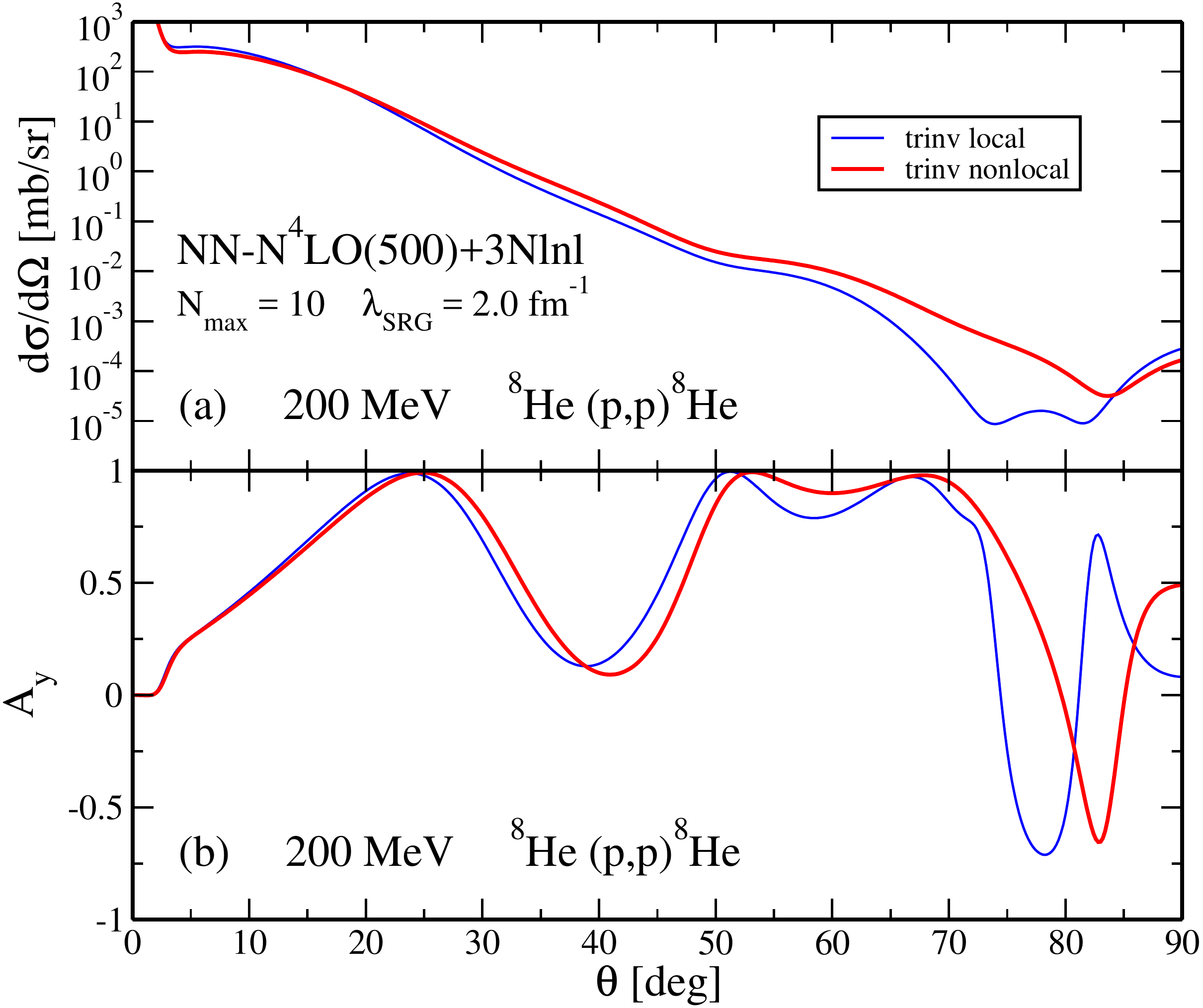}
\caption{\label{8He_200MeV} (Color online) The same as in Fig.~\ref{8He_71MeV} but at the projectile energy of $200$ MeV/nucleon.}
\end{center}
\end{figure}

We now present the results for proton elastic scattering off \textsuperscript{6}He and \textsuperscript{8}He. These nuclei are characterized by the presence of a neutron
halo surrounding an $\alpha$ core. Due to this exotic structure, an adequate description of such systems requires more sophisticated approaches and many theoretical papers
have been devoted to this~\cite{PhysRevC.9.1730,PhysRevC.28.364,ZHUKOV1993151,PhysRevC.48.165,PhysRevC.50.189,PhysRevC.55.688,PhysRevC.56.1720,
DANILIN1998383,PhysRevC.61.044601,PhysRevC.61.024319,Fedorov2003,navratil2004translationally,RevModPhys.76.215,PhysRevC.73.021302,
KARATAGLIDIS200740,CRESPO200626,PhysRevC.76.054607,PhysRevC.78.017306,Bacca2009,PhysRevC.82.054002,Ibraeva2011,PhysRevC.86.044601,PhysRevC.86.034321,
PhysRevC.85.044617,orazbayev2013,PhysRevC.88.034610,PhysRevC.90.044004,PhysRevLett.113.032503,PhysRevLett.117.222501,Quaglioni:2017vpa}.
In this current work we use the model for the optical potential, based on the impulse approximation, to test the new nonlocal densities in the description of the differential cross
section. For the sake of completeness we also show the results for the analyzing power, but, as reported in Refs.~\cite{PhysRevC.61.044601,GUPTA200077}, we do not expect
that the model is able to describe the experimental data for this observable. In this regard, improved results could be obtained introducing medium contributions of the residual nucleus
on the the struck target nucleon, and adopting the cluster description of Refs.\cite{PhysRevC.85.044617,PhysRevC.88.034610}, developed to include the internal dynamics of halo
nuclei, such as \textsuperscript{6,8}He.

As we did in the previous section, we start our discussion by, in Fig.~\ref{density_profiles_halo}, displaying the neutron and proton density profiles. Once again, the local ones
are obtained from Eq.~(\ref{ftlocaldens}) and the nonlocal profiles are obtained using Eq.~(\ref{density_profile}). All calculations
were performed using the same interaction employed for stable nuclei. The only difference is represented by the $\nmax$ parameter used to define the model space for
the densities, that is equal to $12$ for \textsuperscript{6}He and $10$ for \textsuperscript{8}He.
In this case, the results obtained with the nonlocal densities exhibit a larger extent toward higher values of momentum transfer, as was observed for stable nuclei in
Fig.~\ref{density_profiles}. Moreover, looking at the results for proton profiles, it is interesting to notice how the difference between the two curves, that is maximum for
\textsuperscript{4}He (see Fig.~\ref{density_profiles}), is systematically reduced with the increasing neutron number.

In Fig.~\ref{6He_71MeV} and in Fig.~\ref{8He_71MeV} we present the results for the differential cross section and the analyzing power computed at $71$ MeV for
elastic proton scattering off \textsuperscript{6}He and \textsuperscript{8}He, respectively. The experimental data are taken from Refs.~\cite{Hatano2005,PhysRevC.82.021602,
PhysRevC.84.024604,PhysRevC.87.021601}, where the elastic scattering of \textsuperscript{6}He and \textsuperscript{8}He off a polarized proton target has been measured
at a laboratory kinetic energy of $71$ MeV/nucleon. In both cases the local density seems to give the best description of the cross section, while the results obtained with the
nonlocal density overestimate the experimental data, especially at large angles, beyond the first minimum. As stated above, the energy of the scattering process is too low
and the nuclei under consideration require a more sophisticated description, so it is not surprising that our results cannot reproduce the data for the analyzing power.

As we did in Sec.~\ref{sec_nonlocaldens_results}, we display in Fig.~\ref{8He_conv} the $\nmax$ convergence pattern of the differential cross section for \textsuperscript{8}He,
computed with local and nonlocal densities. In both cases we see that the curves obtained at $\nmax = 0$ overestimate the experimental data and with the increasing value of
this parameter the cross section is pushed down. For the local density, the final result obtained at $\nmax = 10$ is in a good agreement with the data, while for the nonlocal density
it still overestimates the data. We performed the same calculations also for \textsuperscript{4}He and \textsuperscript{6}He finding similar patterns, but in those cases the difference
between the curves is smaller and for \textsuperscript{4}He is not appreciable.

Finally, in Fig.~\ref{6He_200MeV} and in Fig.~\ref{8He_200MeV} we show our predictions for the differential cross section and the analyzing power of \textsuperscript{6}He
and \textsuperscript{8}He computed at $200$ MeV using the same interaction employed in previous calculations. Here it is particularly interesting to notice the different
behaviour displayed by the analyzing power around $50$ degrees, that is negative for \textsuperscript{6}He and positive for \textsuperscript{8}He.

\section{Summary and conclusions}
\label{sec_conclusions}

The purpose of this work was to achieve another step toward a consistent calculation of a microscopic optical potential, using chiral interactions as the only input. We extended the approach of Ref.~\cite{navratil2004translationally} to generate {\it ab initio} translationally-invariant nonlocal one-body densities. These densities were then used to construct a microscopic optical potential for elastic proton-nucleus scattering, that was derived at the first order within the spectator
expansion~\cite{PhysRevC.52.1992,PhysRevC.56.2080,PhysRevC.57.1378} of the nonrelativistic multiple scattering theory and assuming the impulse approximation.
The optical potential was computed performing the folding integral between two basic ingredients: the free $NN$ $t$ matrix and the nuclear density matrix. 
The $t$ matrix was computed using the NN-N\textsuperscript{4}LO(500) chiral interaction~\cite{PhysRevC.91.014002,PhysRevC.96.024004}, while the density matrices
were obtained using the NN-N\textsuperscript{4}LO(500)+3Nlnl chiral interaction, which includes the N$^2$LO three-nucleon force with simultaneous local and nonlocal
regularization~\cite{Navratil2007, Leistenschneider:2017mrr}.

The calculation of the nonlocal densities requires the knowledge of the many-body nuclear wave functions, that in our case were obtained from the {\it ab initio} NCSM approach.
This method employs realistic two- and three-nucleon forces and is particularly suited for the description of light nuclei, because all nucleons are treated as active degrees of freedom. This allows us to include many-nucleon correlations in our calculations that consequently provide high-quality wave functions. 

In Sec.~\ref{sec_nonlocaldens_results} we displayed the plots of the neutron and proton nonlocal densities for \textsuperscript{4,6,8}He, \textsuperscript{12}C,
and \textsuperscript{16}O, obtained using NCSM. All results were shown with and
without the COM removal, in order to better appreciate the difference. As illustrated in the results, the sizeable differences between the {\it wiCOM} and {\it trinv} density motivates
the need for procedural COM removal in light nuclei if accurate density related observables are to be computed. We expect that some observables will amplify the effects observed
by COM removal and thus better gauge the importance of the procedure in heavier nuclei. In addition, the production of a general nonlocal density will allow for proper
{\it ab initio} treatment of densities in density dependent calculations and reduce the number of approximations.

In Sec.~\ref{sec_opresults} we used the nonlocal densities to compute the optical potential and then the $pA$ elastic scattering observables.
We performed our calculations for all nuclei considered in Sec.~\ref{sec_nonlocaldens_results} and for different kinetic energies of the incident proton in the laboratory frame. 
We chose \textsuperscript{4}He as a case study to analyze the differences between the scattering observables obtained with nonlocal and local densities, as well as with {\it trinv} and {\it wiCOM} densities. Our results display a great improvement obtained with the {\it trinv} nonlocal density, compared with those obtained
with the local one, that allowed us to describe the experimental data of the differential cross section up to large values of the scattering angles. Similar conclusions, but
with a smaller difference, were also found for \textsuperscript{12}C and \textsuperscript{16}O. We also tested our densities in the description of the scattering observables
for \textsuperscript{6,8}He, finding also in this case a reasonable agreement with the experimental data of the differential cross section, especially for \textsuperscript{6}He.

In conclusion, these {\it ab initio} translationally invariant nonlocal one-body densities provide a better description of the experimental data for $pA$ elastic scattering, especially
for light systems like \textsuperscript{4}He. Although we found good results for stable nuclei, the model for the optical potential that we used was not good enough to describe
exotic nuclei with halo structure, like \textsuperscript{6,8}He. Future calculations, based on more sophisticated methods, could provide interesting results for such systems,
in particular for the analyzing power, which seems to display different behaviour for \textsuperscript{6}He and \textsuperscript{8}He. Finally, as future improvements, we also plan
to include three-body interactions in the model for the optical potential as well as medium effects to better describe experimental data at lower energies.

\appendix
\section{}
\label{sec_appendix}

Here we present the relationship between the nonlocal translationally invariant density presented here and the local version derived in Ref.~\cite{navratil2004translationally}.  We recall the formula for the translationally invariant nonlocal density (\ref{eqn:trinvnlocdens}),
\begin{equation}
\begin{split}
&\langle A \lambda_j J_j M_j\, \vert \rho_{op}^{trinv}(\vec{r}-\vec{R},\vec{r}\,'-\vec{R}) \vert \, A \lambda_i J_i M_i \rangle \\
& = \Big(\frac{A}{A-1}\Big)^{\frac{3}{2}}\sum \frac{1}{\hat{J}_f} ( J_i M_i K k \vert J_f M_f ) \\
&\quad \times \big(M^K\big)_{nln'l',n_1l_1n_2l_2}^{-1} \, \bigg(Y_l^*(\widehat{\vec{r}-\vec{R}})\,Y_{l'}^*(\widehat{\vec{r}\,'-\vec{R}})\bigg)_k^{(K)} \\
&\quad \times R_{n,l}\Big(\sqrt{\frac{A}{A-1}} \vert \vec{r}-\vec{R} \vert \Big) R_{n',l'}\Big(\sqrt{\frac{A}{A-1}} \vert \vec{r}\,'-\vec{R} \vert \Big) \\ 
&\quad \times (-1)^{l_1+l_2+K+j_2-\frac{1}{2}} \, \hat{j}_1 \hat{j}_2 \left \{
  \begin{tabular}{ccc}
  $j_1$ & $j_2$ & $K$ \\
  $l_2$ & $l_1$ & $\frac{1}{2}$ \\
  \end{tabular}
\right \} \\
&\quad \times {}_{SD} \langle A \lambda_f J_f \vert \vert \,(a_{n_1,l_1,j_1}^{\dagger}\,\tilde{a}_{n_2,l_2,j_2})^{(K)}\,\vert \vert A \lambda_i J_i \rangle_{SD} \,. \\
\end{split}
\end{equation}
To get the local form of the density, we need to evaluate $\rho_{op}^{trinv}(\vec{r}-\vec{R},\vec{r}\,'-\vec{R}) \vert_{\vec{r}=\vec{r}\,'}$.
\begin{equation}
\begin{split}
&\langle A \lambda_j J_j M_j\, \vert \rho_{op}^{trinv}(\vec{r}-\vec{R}) \vert \, A \lambda_i J_i M_i \rangle \\
& = \Big(\frac{A}{A-1}\Big)^{\frac{3}{2}}\sum \frac{1}{\hat{J}_f} ( J_i M_i K k \vert J_f M_f ) \\
&\quad \times \big(M^K\big)_{nln'l',n_1l_1n_2l_2}^{-1} \, \bigg(Y_l^*(\widehat{\vec{r}-\vec{R}})\,Y_{l'}^*(\widehat{\vec{r}-\vec{R}})\bigg)_k^{(K)} \\
&\quad \times R_{n,l}\Big(\sqrt{\frac{A}{A-1}} \vert \vec{r}-\vec{R} \vert \Big) R_{n',l'}\Big(\sqrt{\frac{A}{A-1}} \vert \vec{r}-\vec{R} \vert \Big)  \\ 
&\quad \times (-1)^{l_1+l_2+K+j_2-\frac{1}{2}} \, \hat{j}_1 \hat{j}_2 \left \{
  \begin{tabular}{ccc}
  $j_1$ & $j_2$ & $K$ \\
  $l_2$ & $l_1$ & $\frac{1}{2}$ \\
  \end{tabular}
\right \} \\
&\quad \times {}_{SD} \langle A \lambda_f J_f \vert \vert \,(a_{n_1,l_1,j_1}^{\dagger}\,\tilde{a}_{n_2,l_2,j_2})^{(K)}\,\vert \vert A \lambda_i J_i \rangle_{SD} \,. \\
\end{split}
\end{equation}
We then uncouple the spherical harmonics and make use of
\begin{equation}\label{eq:spherical_harmonics}
\begin{split}
&\bigg(Y_l^*(\hat{r})\,Y_{l'}^*(\hat{r}) \bigg)_k^{(K)}  \\
&\quad = \sum_{K,k} \sqrt{\frac{(2l+1)(2l'+1)}{4\pi(2K+1)}} (l\,0\,l'\,0 \vert K0) \, Y_{Kk}^*(\hat{r}) \, \\
\end{split}
\end{equation}
to arrive at our result. However, in order to match Ref. \cite{navratil2004translationally} we use the following relation for the matrix elements of a spherical harmonic from Ref. \cite{varshalovich1988quantum},
\begin{equation}\label{eq:spherical_harmonics_me}
\begin{split}
&\langle l_1 \frac{1}{2} j_1 \vert \vert Y_{K} \vert \vert l_2 \frac{1}{2} j_2 \rangle = \frac{1}{\sqrt{4\pi}}\,\hat{j}_1 \hat{j}_2 \hat{l}_1 \hat{l}_2 \, (-1)^{j_2+\frac{1}{2}} \\
&\qquad \times (l_1\,0\,l_2\,0 \vert K0) \, \left \{
  \begin{tabular}{ccc}
  $j_1$ & $j_2$ & $K$ \\
  $l_2$ & $l_1$ & $\frac{1}{2}$ \\
  \end{tabular}
\right \} \, . \\
\end{split}
\end{equation}
We substitue (\ref{eq:spherical_harmonics}) and (\ref{eq:spherical_harmonics_me}) into the nonlocal translationally invariant nuclear density to obtain
\begin{equation}
\begin{split}
&\langle A \lambda_j J_j M_j\, \vert \rho_{op}^{trinv}(\vec{r}-\vec{R}) \vert \, A \lambda_i J_i M_i \rangle \\
& = \Big(\frac{A}{A-1}\Big)^{\frac{3}{2}}\sum \frac{1}{\hat{J}_f} ( J_i M_i K k \vert J_f M_f ) \, Y_{Kk}^*(\widehat{\vec{r}-\vec{R}}) \\
&\quad \times R_{n,l}\Big(\sqrt{\frac{A}{A-1}} \vert \vec{r}-\vec{R} \vert \Big) R_{n',l'}\Big(\sqrt{\frac{A}{A-1}} \vert \vec{r}-\vec{R} \vert \Big) \\
&\quad \times (-1)^{K} \, \frac{\hat{l} \hat{l}' (l\,0\,l'\,0 \vert K0)}{\hat{l}_1 \hat{l}_2 (l_1\,0\,l_2\,0 \vert K0)} \big(M^K\big)_{nln'l',n_1l_1n_2l_2}^{-1} \\
&\quad \times \langle l_1 \frac{1}{2} j_1 \vert \vert Y_{K} \vert \vert l_2 \frac{1}{2} j_2 \rangle \\
&\quad \times \frac{-1}{\hat{K}}{}_{SD} \langle A \lambda_f J_f \vert \vert \,(a_{n_1,l_1,j_1}^{\dagger}\,\tilde{a}_{n_2,l_2,j_2})^{(K)}\,\vert \vert A \lambda_i J_i \rangle_{SD} \, , \\
\end{split}
\end{equation}
which is the expected form of the local density derived in Ref. \cite{navratil2004translationally}.

\acknowledgments

This work was supported by the NSERC Grant No. SAPIN-2016-00033. TRIUMF receives federal funding via a contribution agreement with the National Research Council of Canada. Computing support came from an INCITE Award on the Titan supercomputer of the Oak Ridge Leadership Computing Facility (OLCF) at ORNL, from Calcul Quebec, Westgrid and Compute Canada, and from the LLNL institutional Computing Grand Challenge Program.



%

\end{document}